\documentclass[12pt,authoryear]{elsarticle}

\journal{arXiv}
 
\usepackage{amsmath} 
\usepackage{amsfonts} 

\usepackage{hyperref}
 
 \graphicspath{{./figures_arxiv/}}
 
\begin{document}
 
\begin{frontmatter}
 
 \title{Online functional connectivity analysis of large all-to-all networks}
 \author[1,2]{Lorenz Esch}
 \author[1]{Jinlong Dong}
 \author[2,3,4]{Matti Hämäläinen}
 \author[5,1]{Daniel Baumgarten}
 \author[1]{Jens Haueisen}
 \author[5]{Johannes Vorwerk\corref{cor1}}
 \ead{johannes.vorwerk@umit-tirol.at}
 
  \cortext[cor1]{Corresponding author}
 
 \affiliation[1]{organization={Institute of Biomedical Engineering and Informatics, Technische Universität Ilmenau}, city={Ilmenau}, country={Germany}}
     \affiliation[2]{organization={Athinoula A. Martinos Center for Biomedical Imaging, Massachusetts General Hospital, Harvard Medical School}, city={Boston, MA}, country={USA}}
  \affiliation[3]{organization={Department of Radiology, Massachusetts General Hospital, Harvard Medical School}, city={Boston, MA}, country={USA}}
    \affiliation[4]{organization={Department of Neuroscience and Biomedical Engineering, Aalto University School of Schience}, city={Espoo}, country={Finland}}
      \affiliation[5]{organization={Institute of Electrical and Biomedical Engineering, UMIT TIROL - Private University for Health Sciences and Health Technology}, city={Hall in Tirol}, country={Austria}}
 
 \begin{abstract}
 The analysis of EEG/MEG functional connectivity has become an important tool in neural research. Especially the high time resolution of EEG/MEG enables important insight into the functioning of the human brain.  To date, functional connectivity is commonly estimated offline, i.e., after the conclusion of the experiment. However, online computation of functional connectivity has the potential to enable unique experimental paradigms. For example, changes of functional connectivity due to learning processes could be tracked in real time and the experiment be adjusted based on these observations. Furthermore, the connectivity estimates can be used for neurofeedback applications or the instantaneous inspection of measurement results. In this study, we present the implementation and evaluation of online sensor and source space functional connectivity estimation in the open-source software MNE Scan. Online capable implementations of several functional connectivity metrics were established in the Connectivity library within MNE-CPP and made available as a plugin in MNE Scan. Online capability was achieved by enforcing multithreading and high efficiency for all computations, so that repeated computations were avoided wherever possible, which allows for a major speed-up in the case of overlapping intervalls.  We present comprehensive performance evaluations of these implementations proving the online capability for the computation of large all-to-all functional connectivity networks. As a proof of principle, we demonstrate the feasibility of online functional connectivity estimation in the evaluation of somatosensory evoked brain activity.
 \end{abstract}



\end{frontmatter}

\section{Introduction} 

To understand information processing in the human brain, identifying the active brain areas and discovering how these areas are connected and synchronized in response to an external stimulus is highly desirable. Due to their unique time resolution in the millisecond range, electroencephalography (EEG) and magnetoencephalography (MEG) are non-invasive methods that are excellently suited for such investigations. Whereas the active brain areas can be identified by means of EEG/MEG source analysis, the estimation of functional connectivity in source or sensor space enables investigating the interplay between these brain areas. Functional connectivity exclusively relies on analyzing the statistical interdependency of activity in different brain regions without any prior assumptions about structural or causal relations between these regions. It is excellently suited to observing transient and dynamic connectivity processes.

In practice, functional connectivity is represented in form of networks consisting of nodes and (weighted or binary) edges. The nodes can be defined as the measurement sensors or as source locations in the brain volume. The edge weights describe the either directed or undirected strength of the connectivity between the nodes and can be computed once connectivity metrics have been established. To date, a multitude of connectivity metrics have been proposed, which differ significantly both with regard to accuracy and stability but also computational effort \citep{bastos2016tutorial, sakkalis2011review}. The analysis of the obtained connectivity networks allows to describe dynamic changes in brain networks which can, e.g., be a result of different experimental conditions or training effects. Important tools in the analysis of these networks are methods and measures from graph theory, such as the computation of node degree, node strength, clusters, motifs, paths, and hubs \citep{sporns2010networks, stam2004functional, richiardi2013machine, bullmore2009complex}.

A pitfall in the analysis of functional connectivity using EEG/MEG in sensor space is the occurrence of spurious connectivity, though the susceptibility to spurious connectivity differs strongly between connectivity measures. Common phenomena giving rise to spurious connectivity are the use of a common reference average in EEG, where the signal measured by the reference electrode can lead to spurious connectivity, or the effects of volume conduction, as the signal of any individual source is instantaneously picked up by all sensors (with varying magnitude). These problems can be mitigated by unmixing the sensor signals  through source estimation and conducting functional connectivity analysis in the source space \citep{barzegaran2017functional, gross2001dynamic, palva2018ghost, schoffelen2009source, silva2017effect}. However, it has to be observed that also inaccuracies in the source estimation, e.g., due to simplified head models, can influence functional connectivity estimates \citep{cho2015influence}. Besides the erroneous computation of functional connectivity metrics, also brain activity that cannot be measured with EEG/MEG can influence the interpretation of functional connectivity. 

EEG/MEG functional connectivity has been and is used in a multitude of studies to discover the brain mechanisms underlying diverse brain functions, such as motor learning during task execution \citep{sun2007functional}, language processing \citep{mamashli2019oscillatory, maess2016prediction, gaudet2020functional, hyder2021functional}, and even during resting-state \citep{albert2009resting, brookes2011measuring, colclough2016reliable, de2010temporal, mackintosh2021psychotic}. Correlations between functional connectivity and individual task performance were also found for other cognitive functions, including working memory \citep{hampson2006brain, klados2019impact}, visual perceptual learning \citep{lewis2009learning}, statistical learning \citep{toth2017dynamics,paraskevopoulos2017functional}, and face processing \citep{zhu2011resting}. Functional connectivity networks have also been proposed as biomarkers for early disease prediction and prevention, e.g., in schizophrenic patients \citep{hirvonen2017whole}. 
Modulations in $\beta$-band (18 - 30 Hz) activity are assumed to reflect the preparation and performance of voluntary, passive, and imagined movement as well as tactile stimulation \citep{jurkiewicz2006post, pfurtscheller1996event, byrne2017mean}. 
The brain response to median nerve stimulation at either the right or left wrist has been shown to include processing both in the primary (S1) and secondary (S2) somatosensory cortex as well as in the posterior parietal cortex \citep{hari1999magnetoencephalography}. Median nerve stimulation can be used to study healthy \citep{weisend2007paving, wiesman2017oscillatory, huang2000sources} and abnormal \citep{ren2019abnormal, huang2004meg} sensory processing. In this study, median nerve stimulation data are used to validate the new online functional connectivity pipeline.

The estimation of functional connectivity in such studies is usually performed offline, i.e., after the experiment is concluded. However, scenarios are perceivable where online processing of EEG/MEG measurements, i.e., the low-latency evaluation of data during the measurement, is desirable. A basic example would be ``live'' monitoring of the measurement outcomes to allow early intervention to fix possible problems, adjust the measurement time/number of recorded samples to achieve a certain signal-to-noise ratio (SNR), or to assess whether the experimental paradigm works as anticipated. But also innovative, dynamic experimental designs that directly depend on the measured brain states are conceivable. In clinical care, rapid analysis of normal or abnormal information processing, such as in the case of median nerve stimulation, or a direct, intuitive insight into the patient's brain state could lead to faster diagnosis and treatment.

To date, focii of online EEG/MEG processing have been brain computer interface (BCI) and neurofeedback applications. Besides the development of basic sensor space operations, such as noise reduction and estimation of frequency band power, also methods for the online estimation of cortical activity have been implemented \citep{dinh2015real, dinh2018real,guarnieri2021rt}. Even though the potential of functional connectivity analysis in online EEG applications such as BCI has been evaluated in numerous studies \citep{tabarelli2022functional,feng2020functional,lee2014classifying,shamsi2021early,torres2020eeg}, only a few implementations that allow at least a basic online estimation of EEG functional connectivity have been presented to date. This might be due to the computational complexity of the connectivity methods and/or the fact that the estimation of functional connectivity based on few samples, as it is common in online scenarios, is a complex task. Particularly the computation of all-to-all connectivity, i.e., functional connectivity analysis without restrictive prior assumptions regarding the involved brain areas, in networks with a large number of nodes results in high computational complexity. Therefore, the connectivity estimation is often restricted to a small number of regions of interest (ROIs), or the number of sensors included is reduced in sensor space analysis.

In this paper, we describe the implementation of online functional connectivity estimation in the open-source software package MNE Scan, study the computational performance in large all-to-all connectivity networks of this implementation in scenarios using simulated and realistic data, and demonstrate the online capability of the obtained pipeline.
 
\section{Methods}
\subsection{Connectivity Metrics}
The main novelty presented in this study is the implementation of online functional connectivity estimation for considerably large networks. Nine functional connectivity metrics were implemented, namely correlation (COR) \citep{schoffelen2009source}, cross-correlation (XCOR), coherency (COHY) \citep{bendat2011random}, coherence (COH) \citep{bendat2011random}, imaginary coherence (IMAGCOHY) \citep{nolte2004identifying}, phase-locking value (PLV) \citep{bruna2018phase}, phase-lag index (PLI) \citep{stam2007phase} and its variations USPLI, WPLI, and DSW PLI \citep{vinck2011improved}. The exact definitions of these metrics can be found in \ref{sec:metrics}. The basic implementation of these metrics is described in Section \ref{sub:implementation} and the resulting processing pipeline is described in Section \ref{sub:pipeline}.

Two of the nine metrics are reflecting relationships in the time domain (COR, XCOR), whereas the other seven are spectral metrics based on the frequency domain representations of the signals. Of course, there are considerable differences between these metrics with regard to the computational complexity, which are usually of minor importance in offline data analysis but crucial in an online scenario. The effect of these differences on the computational effort in the online capable implementations of the connectivity metrics will be evaluated in Section \ref{sub:performance}.

Besides the computational effort, the connectivity metrics also considerably differ with regard to the accuracy and stability of the obtained connectivity estimates. For offline analysis, a multitude of studies exploring the advantages and disadvantages of the different metrics have been performed \citep{schoffelen2009source,vinck2011improved,bastos2016tutorial,sakkalis2011review,barzegaran2017functional}. The results of offline evaluations of functional connectivity metrics for the most part translate to the case of online connectivity analysis. However, due to the commonly low SNR achieved in online evaluations, the robustness of the metrics towards noise plays a crucial role here. The scenarios for which we evaluated the accuracy and robustness of the different functional connectivity metrics are described in Section \ref{sub:evaluation} and the results are presented in Section \ref{sub:evaluation-res}.

 For reasons of conciseness, we will only display the results for the metrics COH, IMAGCOH, PLI, and XCOR within the main manuscript, which stand exemplary for the three main types of metrics considered (time domain, spectral taking phase of the signal into account, spectral using imaginary part). The results for all metrics can be found in \ref{sec:supp-figures}.

\subsection{MNE-CPP}
\label{sub:mne-cpp}
The MNE-CPP project provides a framework to develop and build online as well as offline analysis software for electrophysiological data. The core development is done solely in C++ (CPP) and is therefore targeting more experienced programmers, whereas the resulting stand-alone applications can be easily used by researchers and clinical personnel without coding background. MNE-CPP is based on a two-layer architecture. The library layer provides the core functionalities, which can be used for the development of stand-alone applications and is loaded during runtime. The stand-alone applications can either be realized with a graphical user interface (GUI) or a command line interface (CLI). Furthermore, a multitude of examples and tests are implemented on the application layer of MNE-CPP.

To maximize cross-platform compatibility and simplify the project maintenance, MNE-CPP makes use of only one external dependency, namely the Qt framework (\url{https://www.qt.io}). Furthermore, it uses the Eigen library via a so-called ``clone-and-own'' approach (\url{https://eigen.tuxfamily.org}). Qt provides tools for GUI creation whereas Eigen provides tools for linear algebra computations. Minimizing the use of software of unknown provenance (SOUP) is favorable when developing medical software applications that have to meet regulatory requirements: each third-party dependency must be tracked, and its development life cycle must be auditable. Furthermore, all dependencies are able to compile on multiple platforms and devices for cross-platform capability in order to develop stand-alone applications on Windows, macOS, and Linux \citep{esch2018mne}. MNE-CPP is open-source BSD licensed (clause 3) and accessible via \url{https://mne-cpp.github.io/} and \url{https://github.com/mne-tools/mne-cpp}.

Further details regarding the architecture of MNE-CPP can be found in the reference publication \citep{esch2019mne} and on \url{https://mne-cpp.github.io/}.

\subsection{MNE Scan}
MNE Scan is a tool developed within the MNE-CPP software project to acquire data from MEG/EEG devices and process the resulting data streams online. MNE Scan is designed as a plug-in-based software, meaning that acquisition and processing tasks are developed as individual units. This ensures a modular software architecture, which can easily be extended by new plug-ins with support for new hardware devices or analysis methods. MNE Scan accepts two kinds of plug-ins: Algorithm and Sensor plug-ins. Sensor plug-ins can only have output connectors, whereas Algorithm plug-ins can have both in- and output connectors. The workflow in MNE Scan follows a pipeline approach where the user can select and connect Sensor and subsequent Algorithm plug-ins.

Currently, there are nine Sensor plug-ins, which offer connections to MEG (MEGIN, BabyMEG) and EEG (TMSI Refa, EEGoSports, gTecUSB, Natus, BrainAmp, and Lab Streaming Layer/LSL) devices. It is also possible to stream pre-recorded data from disk in order to mimic a measurement session and test the online processing. The available Algorithm plug-ins include online capable implementations of, e.g., temporal filtering, signal space projection (SSP) \citep{uusitalo1997signal}, SPHARA \citep{graichen2015sphara}, trial averaging, source estimation, and - as presented in this study - functional connectivity estimation.

Further details regarding the architecture of MNE Scan can be found in the reference publication \citep{esch2018mne} and on \url{https://mne-cpp.github.io/pages/documentation/scan.html}.

\subsection{Implementation of functional connectivity metrics}
\label{sub:implementation}

The functional connectivity metrics were implemented in the new Connectivity library as part of the MNE-CPP library layer. This library includes routines to calculate functional connectivity metrics, a container to store the resulting networks, and an Application Programming Interface (API). All features were implemented with the goal of achieving a performance that enables the online capability of the resulting applications.

\begin{figure}[htb]
\begin{center}
\includegraphics[width=0.7\textwidth]{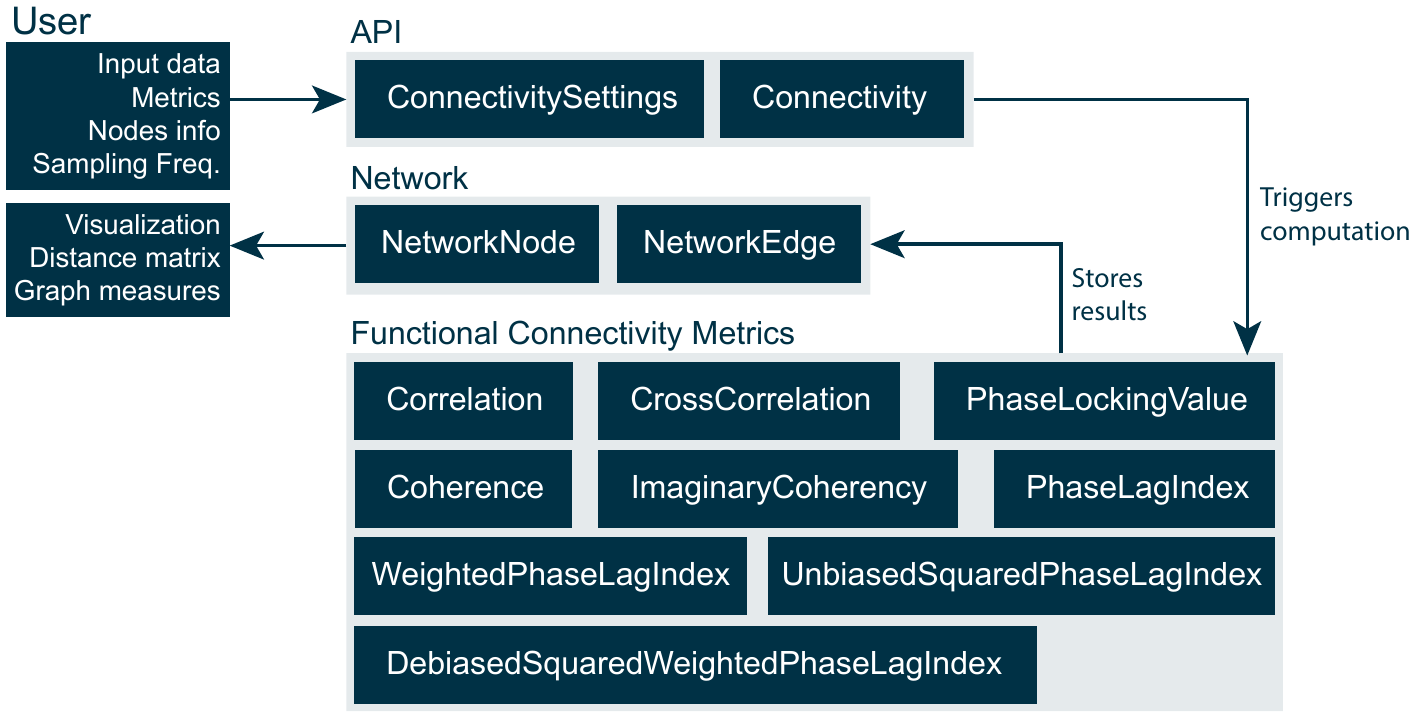}
\end{center}

\caption{The three components of MNE-CPP’s connectivity library. The Connectivity class can be used to trigger the actual computations provided by the functional connectivity metric classes. The final result is provided back to the user in form of a Network class, which can then be used to analyze the network with the help of visualization tools, graph measures, etc.}
\label{fig:class}
\end{figure}

The library was designed to be as generic as possible in order to not limit its functionality to only a set of specific use cases or measurement modalities. It should further be extensible to include new connectivity metrics in the future. The library can be divided into three groups of classes, namely the API classes, the actual computational classes for each functional connectivity metric, and the network (data) container classes (see Figure \ref{fig:class}). Here, connectivity networks are computed over trials rather than over time. Connectivity resolved over trials is favorable when dealing with evoked responses, whereas instantaneous connectivity estimates per sample via time-resolved connectivity might be preferable when dealing with spontaneous and resting-state data. However, instantaneous connectivity is not explicitly supported by the current implementation of the library, but spontaneous or resting-state data can be processed by the library in form of data blocks and handled as if they were multiple trials. Such a workflow is also common in resting-state functional connectivity studies.

In the implementation of the functional connectivity metrics, special emphasis was given to three key aspects to ensure the efficient implementation of the Connectivity library. First, multithreading was employed to process each trial in parallel. Second, the parallel functions were optimized with respect to efficient usage of iterative steps in loops, data storing, etc. Third, reusable data are carried over in order to ease the computational burden for subsequent iterations.

The COR metric is realized as a simple dot product and division by the number of samples. The XCOR metric is utilizing an FFT convolution, which consists of the creation of the signal spectra, multiplication of these spectra, and the transformation of the product back to the time domain. Subsequently, the highest correlation value and its index are returned.

All phase synchronization based metrics are calculated in a very similar manner exploiting a running average to avoid unnecessary repetitions of computations, as we will explain in the example of the COH metric. For each trial, the frequency spectrum, cross-spectral density (CSD), and power spectrum density (PSD) are calculated. If these values have been calculated in a previous step for this trial, the computation will be skipped and the stored values will be used. Besides the CSD and PSD per trial, also the sums of CSD and PSD over all trials are directly computed, which saves one iteration step for computing the average over all trials after the computation of CSD and PSD for all trials is finished. Subsequently, these sums are used to perform the actual computation of the COH metric, where the sensors and their corresponding contributions are processed in parallel. Since COH is an undirected metric, only one half of the edge weights needs to be calculated. This procedure is applied for all implementations of the phase synchronization based metrics, adapted to the respective definition of each metric.

The numerous necessary FFT computations  are realized using efficient external libraries, which are distributed with the Eigen library, so that this does not add any additional external dependencies. The Connectivity library's FFT backend supports the Keep It Simple Stupid Fast Fourier Transform (KISSFFT, \url{https://github.com/mborgerding/kissfft}) and Fastest Fourier Transform in the West (FFTW, \url{http://www.fftw.org}).

Further details regarding the underlying architecture and the implementation of the Connectivity library can be found on \url{https://mne-cpp.github.io/pages/development/api\_connectivity.html}.

\subsection{Online functional connectivity pipeline}
\label{sub:pipeline}

\begin{figure}[htb]

\begin{center}
\includegraphics[width=0.7\textwidth]{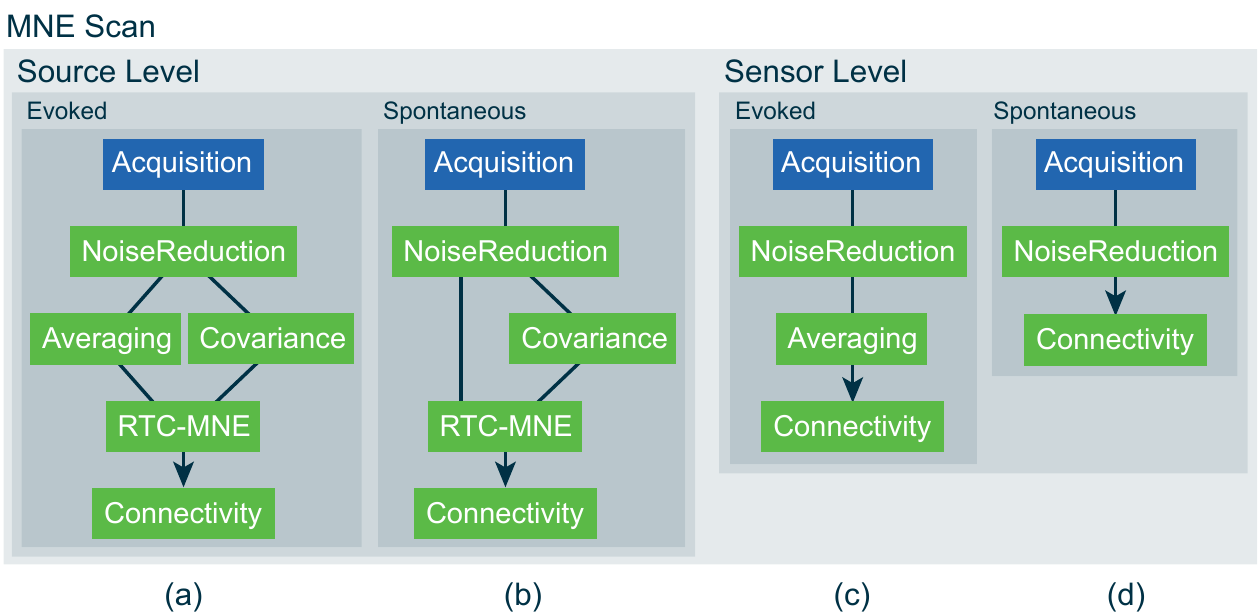}
\end{center}

\caption{Examples of the implemented online functional connectivity pipelines in MNE Scan. Four different modes can process source and sensor level data from measurements of either evoked response experiments or spontaneous activity. The sensor (acquisition) plug-in is depicted in blue whereas the algorithm (processing) plug-ins are shown in green.}
\label{fig:pipeline}
\end{figure}

The functionalities of the connectivity library in combination with newly implemented tools for visualization, the Disp3D library, were used to implement a new online connectivity pipeline in MNE Scan. The plug-ins included in the pipeline depend on the scenario being investigated. Four possible modes reflecting different scenarios are depicted in Figure \ref{fig:pipeline}.

Functional connectivity networks can be estimated and visualized based on sensor- and source-level data. The computation can be done with spontaneous or averaged data as obtained from evoked response experiments.

The possible choices for the acquisition plug-ins correspond to the sensor plug-ins described in Section \ref{sub:mne-cpp}; the implementation of additional sensor plug-ins based on a self-development kit (SDK) provided by the respective manufacturer is commonly straightforward following the architecture of the existing sensor plug-ins.

The Noise Reduction plug-in is implemented with the goal of also performing the necessary data preprocessing within online evaluation pipelines. It includes temporal filtering, SSP and synthetic gradiometers \citep{sun2018noise}, and SPHARA methods \citep{graichen2015sphara} and has been previously described in \citet{esch2018mne}.

For temporal filtering, a design tool can be used to create finite impulse response (FIR) filters on the fly during the ongoing measurement. The design process is controlled by the cutoff frequencies, transition bandwidth, filter method, and filter taps. Low-, high-, and band-pass filters are supported. All implemented filters provide a linear phase response and hence a constant group delay. This property is particularly important when dealing with functional connectivity metrics based on phase synchronization. The actual filter operation is realized with an FFT convolution, where the signal's frequency representation is multiplied by the filter's frequency response. The implemented FFT backends have been described before. All channels are processed in parallel. Since the data are received in form of blocks, the overlap-add method was implemented to cope with the delay introduced by filtering.

The Averaging plug-in processes trigger channels present in the incoming data stream in order to detect events \citep{esch2018mne, dinh2015real}. The distinction between multiple event types is supported to allow for experiments with more than one stimulus type. The user can adjust the start and endpoint of the trial in relation to the trigger. The trials are then cut out of the incoming data stream according to the detected event types. If necessary, a moving average is employed to average the trials and produce the resulting evoked responses for each event type and make them available for subsequent connected plug-ins.

The Covariance plug-in computes the covariance matrix every time the amount of data specified by the user has been gathered. The resulting covariance matrix is forwarded to the Real-Time Clustered Minimum Norm Estimate (RTC-MNE) plug-in. This process copes with changing noise levels over the course of the measurement session. \citet{dinh2015real} propose a source space reduction through a region-wise clustering based on the Destrieux brain atlas \citep{destrieux2010automatic}, which is provided by the Freesurfer toolbox. Based on the clustered forward solution the inverse operator is calculated every time a new covariance estimate arrives and is then used to map the sensor level data to the source space for each data block.
The Covariance and RTC-MNE plug-ins were already implemented in prior work \citep{dinh2015real}. However, the RTC-MNE plug-in was improved in order to also handle raw data streams (non-averaged data), enabling the analysis of spontaneous data.

Finally, the Connectivity plug-in provides access to the functional connectivity algorithms implemented in the Connectivity library. To allow for a variety of different online functional connectivity pipelines, it accepts three different types of input data. First, spontaneous data, e.g., streamed directly from an acquisition plug-in. Second, evoked response data forwarded by the Averaging plug-in. Third, source-level data, e.g., generated by the RTC-MNE plug-in. In the case of evoked data, the Averaging plug-in is only used to cut out the data segment of interest. This means if functional connectivity based on an evoked response experiment is to be investigated, the number of trials in the Averaging plug-in is always set to one. The actual averaging is performed in the Connectivity plug-in after the functional connectivity estimate has been calculated for each cut out trial. The user can change the number of trials, which the Connectivity plug-in uses to compute a connectivity average. The resulting network is sent to the online visualization display and all subsequent connected plug-ins.

\subsection{Evaluation}
\label{sub:evaluation}

Two different scenarios were used for evaluation, one based on simulated and another one based on actual measurement data. The computations were performed on a Linux CentOS 7.6 desktop setup including an Intel Xeon X5660 CPU with 12 physical (24 logical) cores running at 2.8 GHz, 47 GB RAM, and an NVIDIA Quadro 4000 graphics card.

For an initial verification, each spectral functional connectivity metric was tested against its implementation in the MNE-Python toolbox \citep{gramfort2013meg}. COR and XCOR were tested against their Matlab equivalents. The routines of MNE-CPP, MNE-Python, and Matlab were used to process the same data and compare the final results. No differences were found.

\subsubsection{Test scenario - Simulation}
\begin{figure}

\begin{center}
\includegraphics[width=0.7\textwidth]{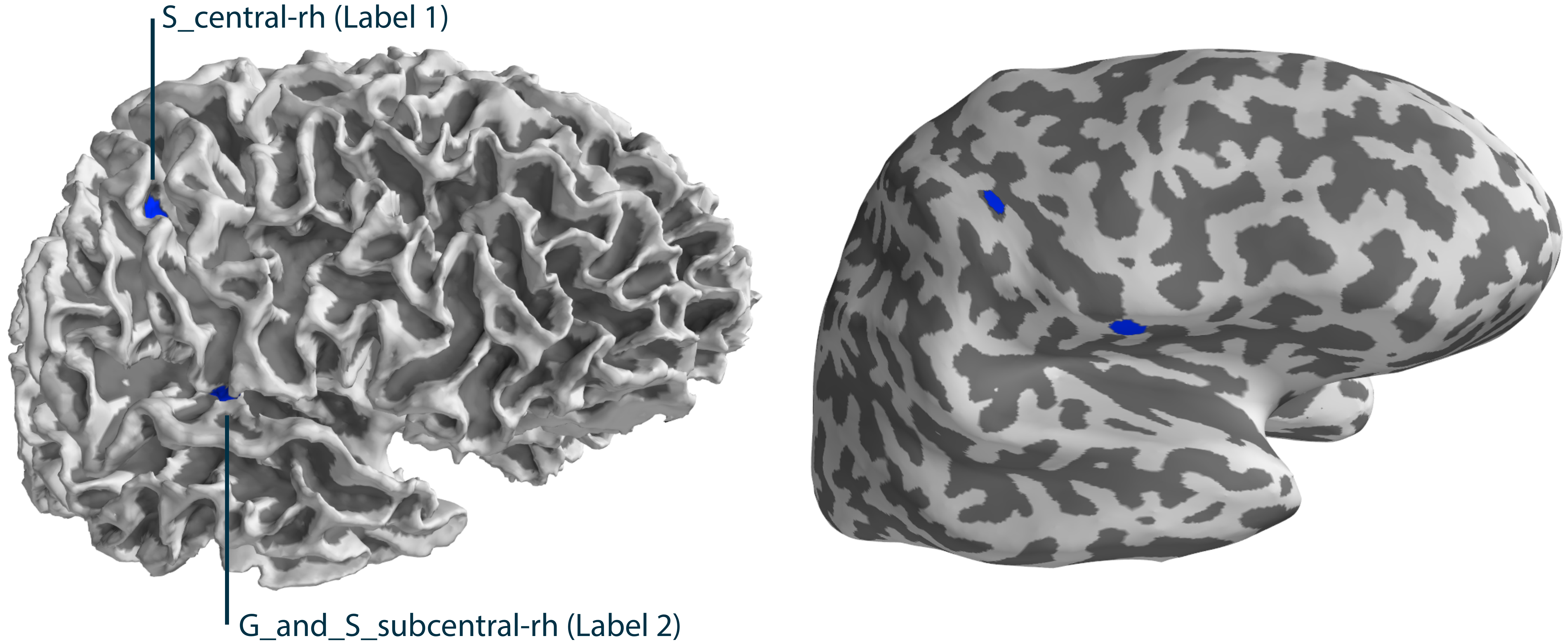}
\end{center}

\caption{Labels defined from the S\_central-rh and G\_and\_S\_subcentral-rh patches.}
\label{fig:scenario1}
\end{figure}

In a first test scenario, a simulation was performed so that the ground truth for source locations and connectivity are known and can be compared with the reconstructions. The simulation was created with the MNE-Python toolbox. Two labels were defined with a 5 mm diameter inside the \mbox{S\_central-rh} and \mbox{G\_and\_S\_subcentral-rh} patches extracted from the Destrieux brain atlas 
 (Figure \ref{fig:scenario1}). Label 1 included two source dipoles which were both driven with an 18 Hz sinusoid at 10 nAm. Label 2 included four source dipoles which were driven with an 18 Hz cosine at 10 nAm each. Based on an existing three-layer BEM forward solution the simulated source signals were projected to MEG sensors. The used forward solution was extracted from the sample\_audvis-eeg-oct-6-fwd.fif file included in the MNE-Sample-Data-Set (\url{https://mne-cpp.github.io/pages/download/sample\_data.html}). Gaussian noise was added to the signals resulting in an SNR of 11.85 dB averaged over all MEG channels. For the source space connectivity estimation, RTC-MNE was chosen as the inverse method based on a reduced source space of 243 sources with the regularization parameter set to $\lambda = 1/SNR^2$ \citep{dinh2015real}. The duration of the simulated signal was set to 160 ms.

\subsubsection{Test scenario - Measurement}
\label{sub:scenario-test}

To simulate a real measurement session, the FiffSimulator plug-in was used to stream data from the pre-recorded MIND data set \citep{weisend2007paving, dataset2022mind}. The paradigm was chosen to be a right-hand median nerve stimulation. A source-level pipeline based on evoked data was chosen, as depicted in Figure \ref{fig:pipeline} (a). The pipeline included a noise reduction step with activated temporal filtering and SSPs. For comparison, the data were analyzed offline as well.

\begin{figure}

\begin{center}
\includegraphics[width=0.7\textwidth]{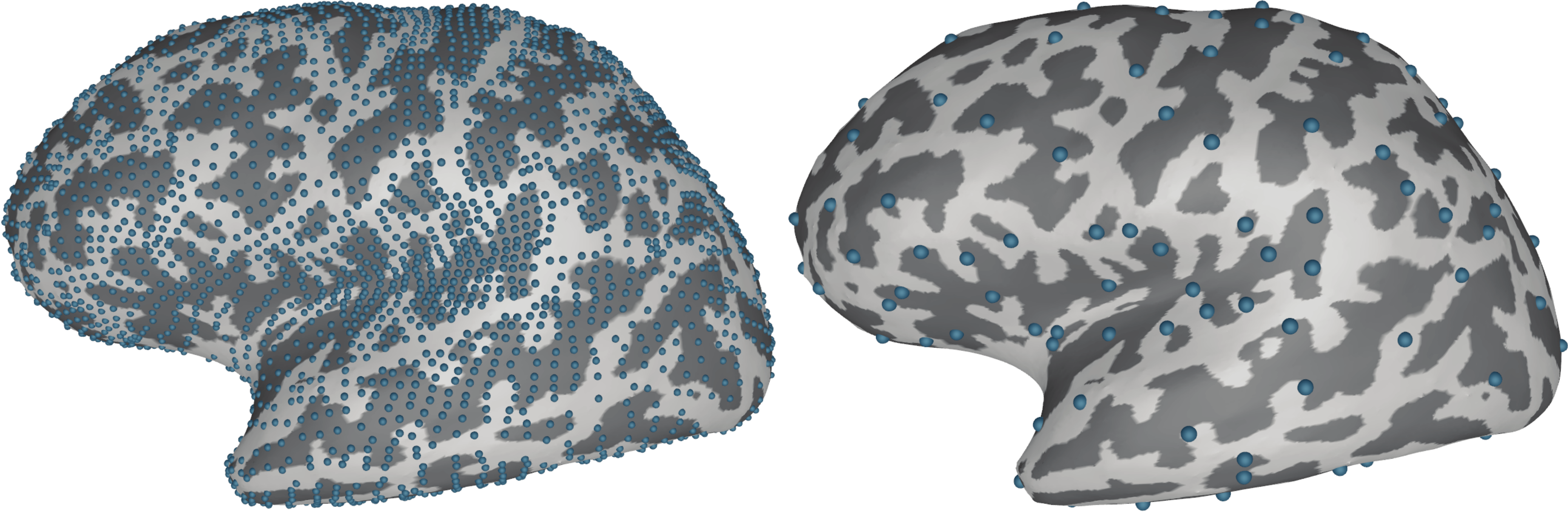}
\end{center}

\caption{The source space for subject \textit{mind010} before and after clustering with the RTC-MNE method based on the \textit{aparc.a2005s} annotation information.}
\label{fig:scenario2}
\end{figure}

Right-hand median nerve stimulations of subject mind010 were extracted and filtered with a 2 Hz cutoff highpass filter. Trials ranging from -100 to 400 ms with an EOG higher than 150 mV, excluding the stimulus artifact, were rejected. This left 200 trials remaining for processing. Each trial was baseline corrected with the pre-stimulus section. The three-layer BEM forward model was clustered based on the Destrieux atlas (aparc.a2005s) \citep{fischl2004automatically}. The cluster number was set to 40 which reduced the source space from 7,423 to 265 active sources. The resulting source space is depicted in Figure \ref{fig:scenario2}.

\section{Results}
 
Besides the evaluation of the accuracy of the functional connectivity estimation in the simulated and realistic data scenario introduced in Section \ref{sub:evaluation}, we also performed an in-depth evaluation of the computational performance of the Connectivity library. The results of this evaluation are presented first of all in this section.
 
\subsection{Performance}
\label{sub:performance}

\begin{figure}

\begin{center}
\includegraphics[width=0.9\textwidth]{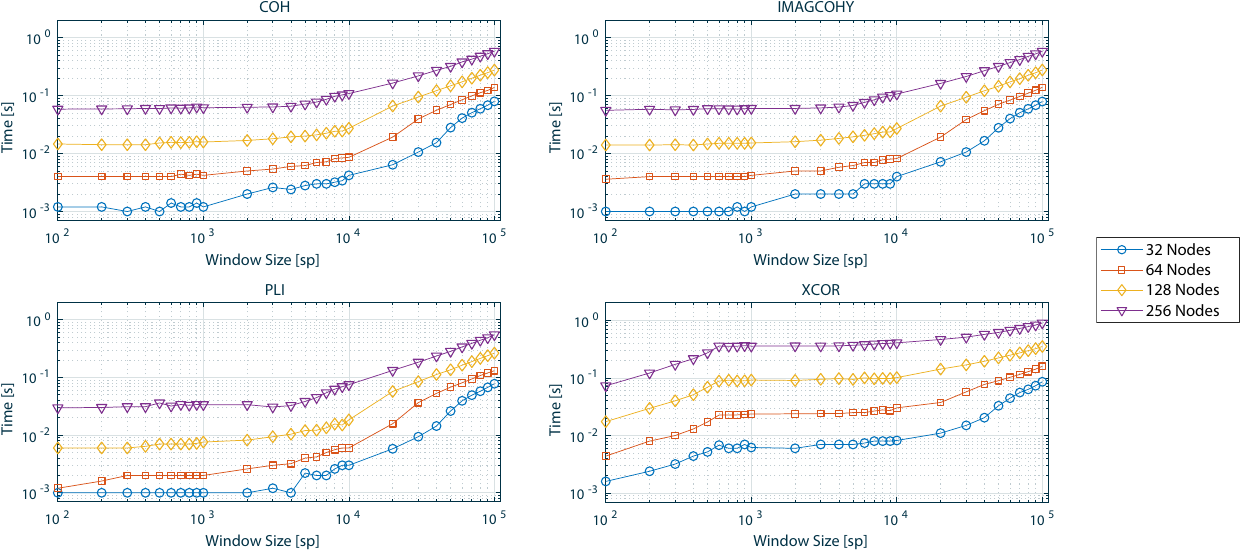}
\end{center}

\caption{Computational timing values in seconds for one trial and different window sizes in samples (sp) as well as number of computed nodes (y-axis in logarithmic scale) for COH, IMAGCOHY, PLI, and XCOR metrics. Computations were generated averaged over five repetitions. Results for all metrics can be found in Figure \ref{fig:supp-timing1}.}
\label{fig:timing1}
\end{figure}

\begin{figure}

\begin{center}
\includegraphics[width=0.9\textwidth]{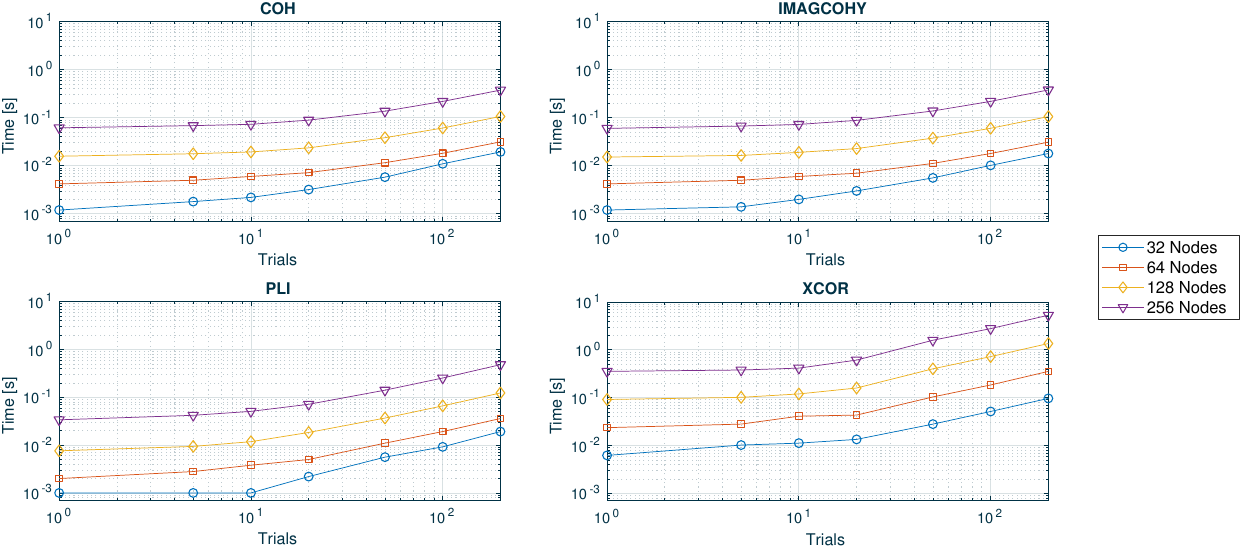}
\end{center}

\caption{Computational timing values in seconds for 1,000 sp and different number of trials as well as computed nodes (y-axis in logarithmic scale) for COH, IMAGCOHY, PLI, and XCOR metrics. Computations were generated averaged over five repetitions. Results for all metrics can be found in Figure \ref{fig:supp-timing2}.}
\label{fig:timing2}
\end{figure}

Each functional connectivity metric was evaluated with different numbers of channels and trials and with different window sizes. Each computation was repeated five times; the performance data reported are averages over the five repetitions. The analysis steps included the actual metric calculation and creation of the final network data container. The FFT length was set to 600 bins, matching the sampling frequency of 600 Hz to result in a frequency resolution of 1 Hz. The FFTW was chosen as the FFT backend and the frequency band was chosen from 8 to 12 Hz (4 bins). The storage mode was turned off, meaning the connectivity estimates for each trial were recalculated for each performance run.

The performance values (computing time) determined for each functional connectivity metric and one trial over various window sizes as well as computed nodes are presented in Figure \ref{fig:timing1}. The performance values determined for each functional connectivity metric and 1,000-sample window size over various numbers of trials as well as computed nodes are presented in Figure \ref{fig:timing2}.

The timing results as a function of the number of samples show similar timing values that are approximately constant for all frequency-based metrics up to 5,000 sp to 6,000 sp (Figure \ref{fig:timing1}), where PLI is less computationally costly than COH and IMAGCOHY. The XCOR timing values are higher compared to every other metric, also when evaluating the timing as a function of the number of trials. The COR metric outperforms all other metrics (Figure \ref{fig:supp-timing1}).

Also when run on multiple trials the frequency-based metrics show similar timing values, again with PLI being slightly more efficient than COH and IMAGCOHY for small numbers of trials (Figure \ref{fig:timing2}, \ref{fig:supp-timing2}). Again, the XCOR timing values are higher than for the frequency-based metrics.

In both test settings, the timing values increased with the number of network nodes for all metrics, where the increase is about linear for small window sizes.

\subsection{Evaluation}
\label{sub:evaluation-res}
\subsubsection{Simulation}

\begin{figure}

\begin{center}
\includegraphics[width=0.7\textwidth]{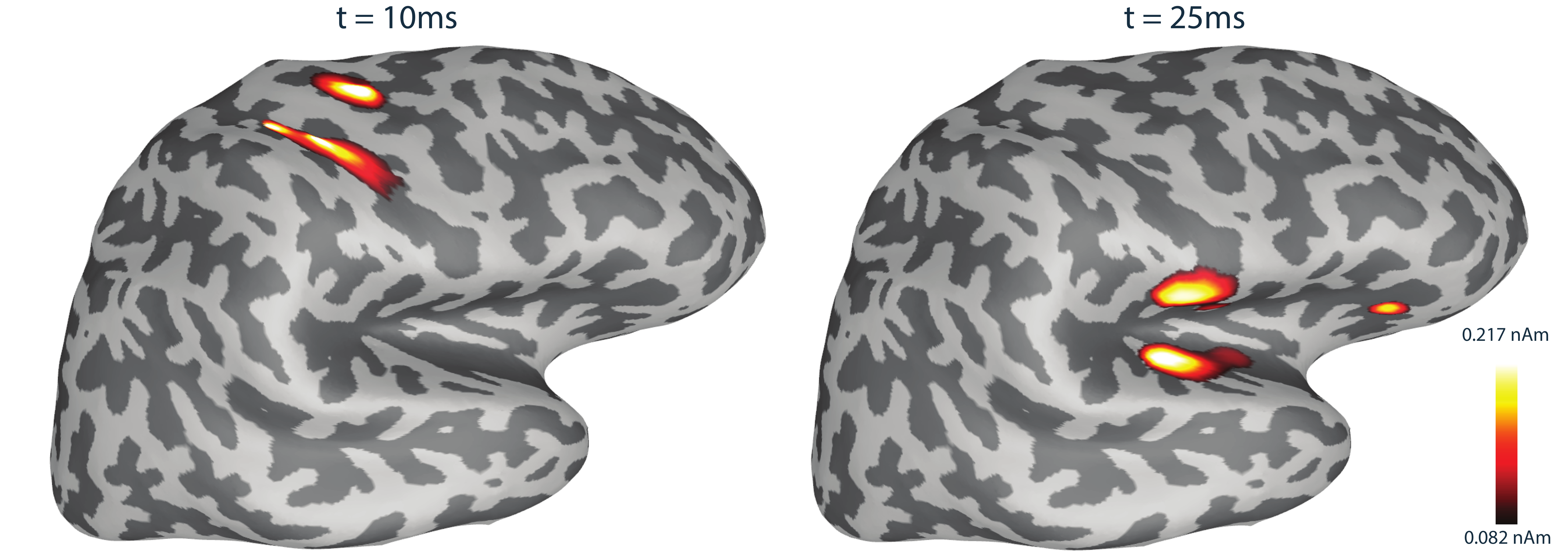}
\end{center}

\caption{Source activity reconstructed with the RTC-MNE method based on simulated data. The processed average included 200 trials. Please note that the absolute value of the source activation was plotted.}
\label{fig:reconstruction-simulation}
\end{figure}

Functional connectivity metrics were computed for the RTC-MNE source reconstructed time courses for 200 trials. In order to be comparable to the steps performed within the pipeline for real data, all trials were baseline corrected with the data segment -50 to 0 ms. Figure \ref{fig:reconstruction-simulation} shows the source reconstruction using the RTC-MNE method at two different time points.

For the connectivity calculation, each trial was cropped from 0 to 160 ms. The sampling rate of the input data was 600 Hz. Frequency bins 0 to 50, which corresponded to 0 to 50 Hz, were averaged to calculate functional connectivity networks. No ROI selection was performed. Instead, all-to-all connectivity networks of the 243 sources were computed. The storage mode of the Connectivity library was turned on, meaning subsequent metric computations used already calculated and stored intermediate data. All networks were normalized based on their maximum edge weight.

\begin{figure}

\begin{center}
\includegraphics[width=0.65\textwidth]{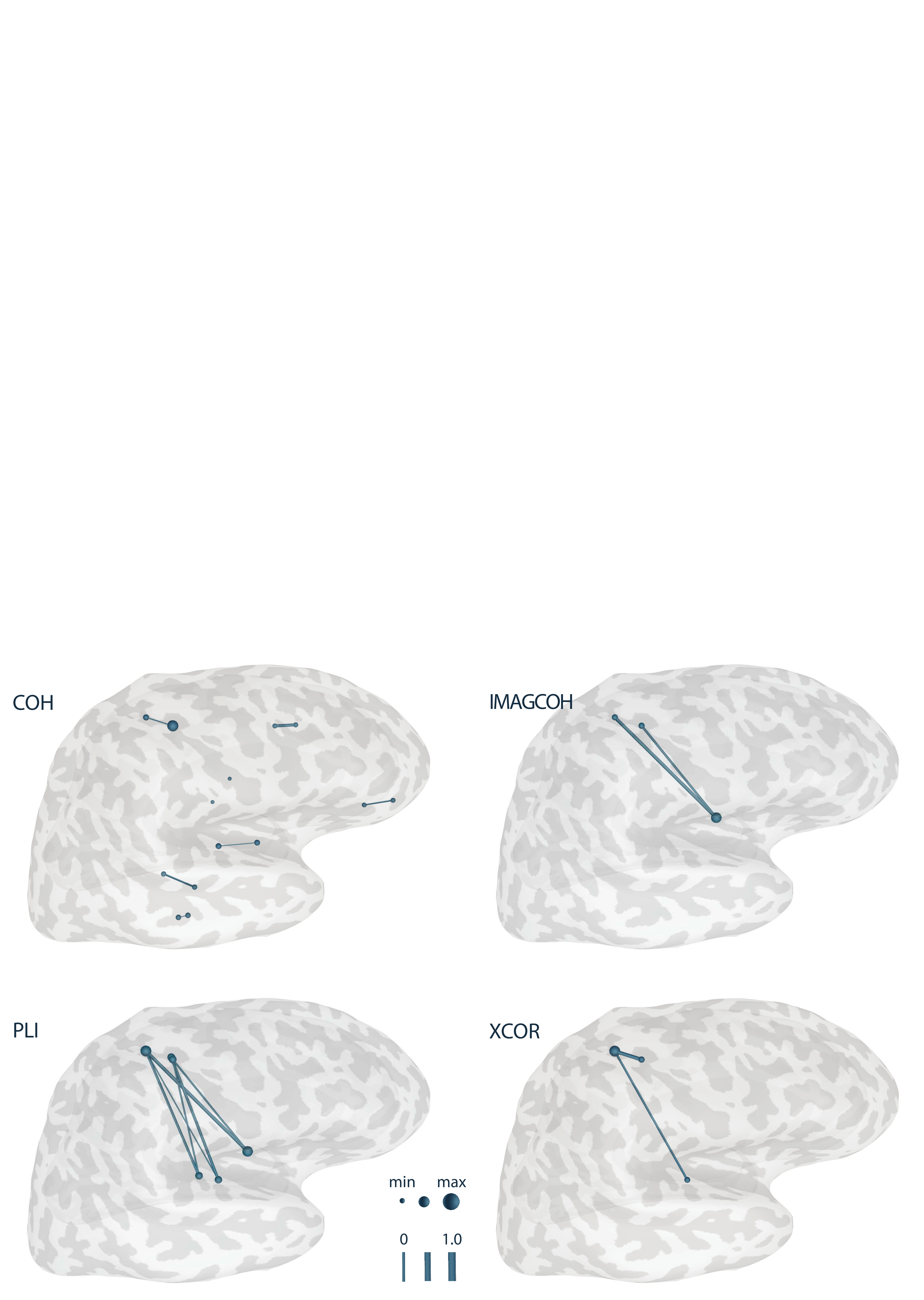}
\end{center}

\caption{Functional connectivity networks for different metrics based on simulated data. The RTC-MNE method was used to compute the source activity. The number of trials was 200. Network nodes are plotted as spheres and edges are represented as tubes connecting the nodes. Edge strength and node degree are represented by their diameter. Please note that the nodes’ sphere diameters are normalized by the node with the maximal value in the thresholded network. Results for all metrics can be found in Figure \ref{fig:supp-connectivity-simulation}.}
\label{fig:connectivity-simulation}
\end{figure}

Figure \ref{fig:connectivity-simulation} shows the visualization of the COH, IMAGCOHY, PLI, and XCOR functional connectivity metrics with the edges representing the strongest 5\% of connections. For each metric and corresponding network, the maximum edge weight was identified and used to normalize all edge weights. The number of trials was 200. It can be observed that the metrics based on the imaginary valued part of the CSD, i.e., IMAGCOHY and PLI, were able to detect the functional connection between the two simulated activation areas L1 and L2. The COH metric, which includes the phase of the signals was not able to detect this functional connection. These trends also hold true for other metrics based on the imaginary part of coherency or including the phase of the signal, respectively, such as WPLI and DSWPLI on the one hand or PLV on the other hand (see Figure \ref{fig:supp-connectivity-simulation}). Also, XCOR seemed to be able to detect some functional connection between L1 and the temporal lobe activity spread from L2.

\begin{figure}

\begin{center}
\includegraphics[width=0.85\textwidth]{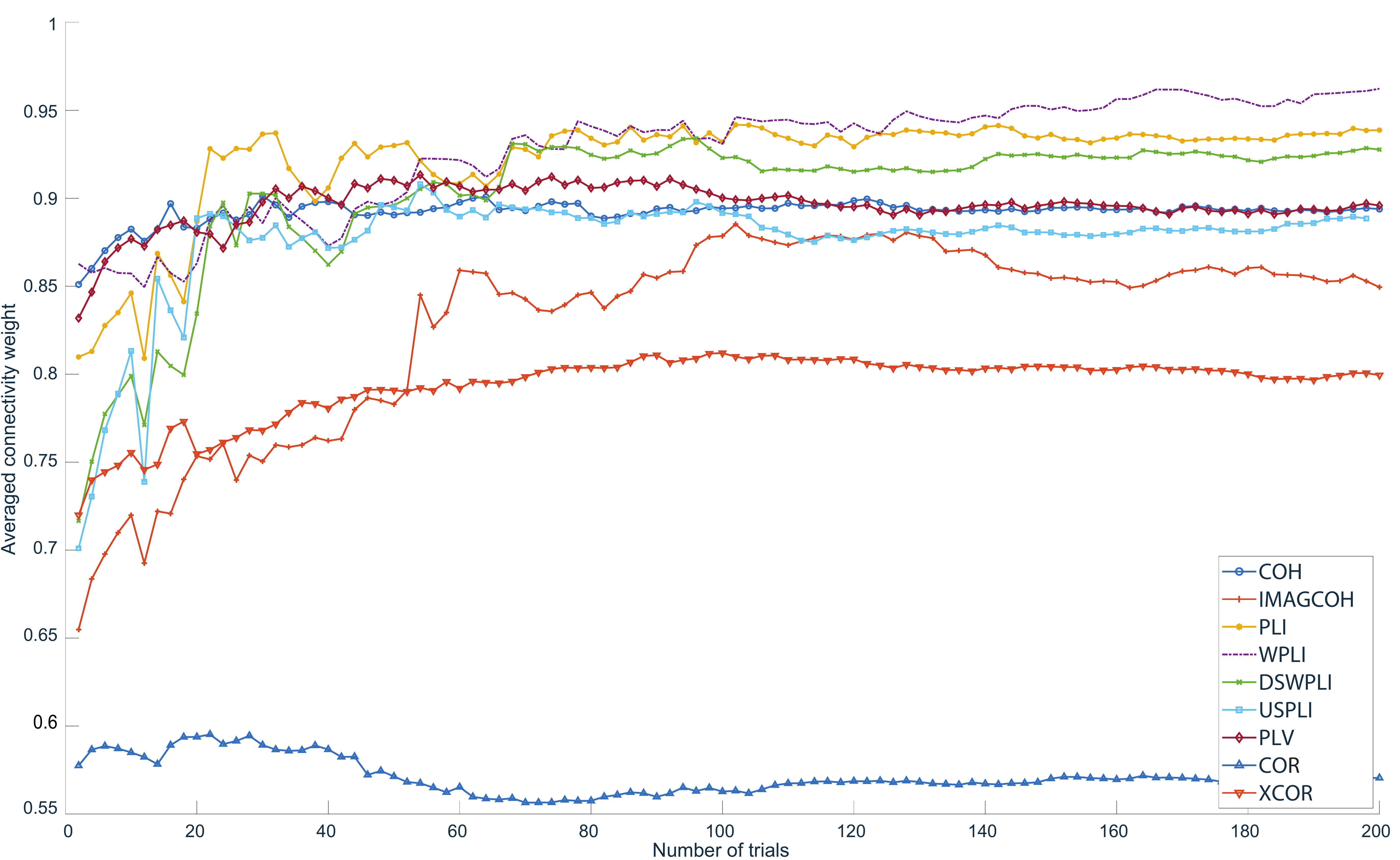}
\end{center}

\caption{The relationship between the final networks at 200 trials and networks calculated at different number of trials. The networks were estimated for all metrics based on simulated data.}
\label{fig:connectivity-simulation-eval}
\end{figure}

The relationship between the connectivity weights and the number of trials is presented in Figure \ref{fig:connectivity-simulation-eval}. For each metric, the 20 strongest edges at 200 trials were identified (Figure \ref{fig:connectivity-simulation}). The exact same edges and their averaged weights were computed for different numbers of trials subsequently. This information gives insight into how the strength of the connections detected in the final network at 200 trials evolved with an increasing number of trials. However, it has to be observed that the convergence of the connection strength with the increasing number of trials does not reflect the correctness of the results; it merely reflects how quickly the detected stable network is established. It can be observed that all metrics except for COR and IMAGCOHY started to stabilize around approximately 40 trials. The averaged IMAGCOHY weights started to stabilize beginning at approximately 60 trials.

\subsubsection{Application/Proof-of-principle}
\label{sub:proof-of-principle}
\begin{figure}

\begin{center}
\includegraphics[width=0.7\textwidth]{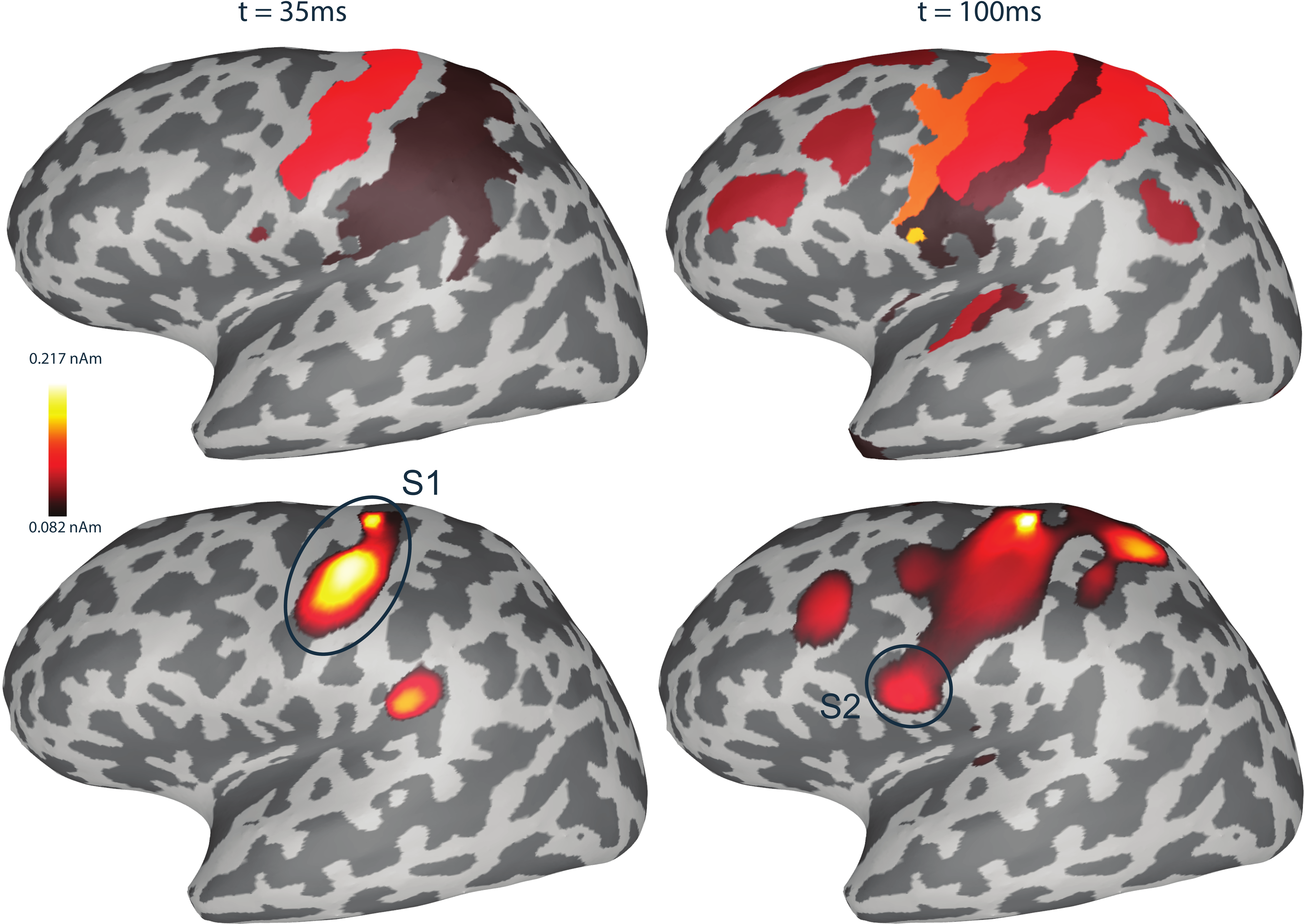}
\end{center}

\caption{RTC-MNE results for a 200 trial average and two time instances based on right-hand median nerve stimulation. The S1 and S2 areas are indicated by blue circles. The upper row presents the patch-based visualization, whereas the lower shows the surface constrained interpolation visualization. Please note that the absolute value of the source activation was plotted.}
\label{fig:reconstruction-experiment}
\end{figure}

\begin{figure}

\begin{center}
\includegraphics[width=0.7\textwidth]{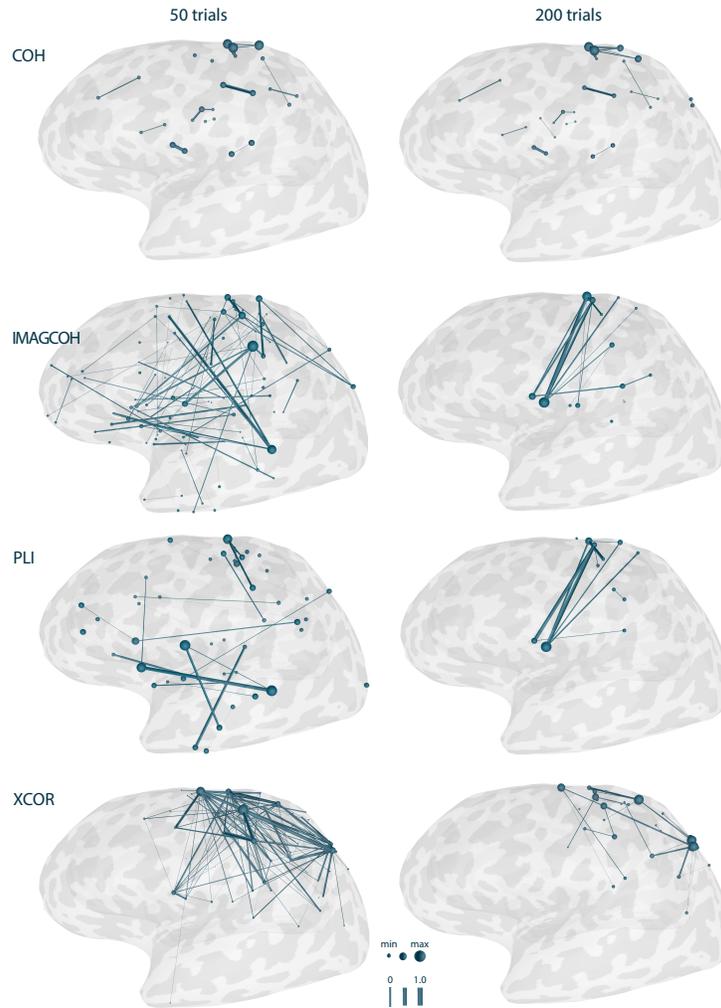}
\end{center}

\caption{Results for functional connectivity metrics implemented in the new Connectivity library based on right-hand median nerve stimulation. Results for 50 and 200 trials are presented. Only the edges representing the strongest 5\% of connections are plotted. Network nodes are plotted as spheres and edges are represented as tubes connecting the nodes. Edge strength and node degree are represented by their diameter. Please note that the nodes' sphere diameters are normalized by the node with the maximal value in the thresholded network. Results for all metrics can be found in Figure \ref{fig:supp-connectivity-experiment}.}
\label{fig:connectivity-experiment}
\end{figure}

RTC-MNE source estimation was performed for the 200 trials obtained as described in Section \ref{sub:scenario-test} (Figures \ref{fig:reconstruction-experiment}). For these 200 trials, all implemented functional connectivity metrics were computed. The trials were cropped to include data from 10 to 150 ms relative to the trigger. This ensured that the stimulation artifact was cut out. No ROIs were specified, meaning all-to-all networks were estimated. The clustered source space included 265 sources, which were specified as the nodes of the network. Thus, a total of 65,536 edges was calculated for each of the 200 trials and then averaged together to form the final network. The Connectivity library's storage mode was turned on, meaning subsequent metric computations used already calculated and stored intermediate data. The sampling rate of the input data was 1,792 Hz. Frequency bins 18 to 30, which corresponded to 18 to 30 Hz ($\beta$-band), were averaged to calculate functional connectivity networks. All networks were normalized and the edges representing the strongest 5\% of connections were plotted. The results for the four exemplary metrics are shown for 50 and 200 trials in Figure \ref{fig:connectivity-experiment}. It can be observed that the metrics based on the imaginary part of the CSD, i.e., IMAGCOHY and PLI, were able to detect the expected functional connection between the two activation areas S1 and S2, however, only for a large number of trials. Phase and amplitude based metrics, such as COH, were not able to achieve the same results. Their connections were already fixed after a lower number of trials and remained the same even as more trials were added. The same is true for the time domain based metric XCOR. The results for all metrics show that this observed trend also holds true for the other metrics (Figure \ref{fig:supp-connectivity-experiment}).

\begin{figure}

\begin{center}
\includegraphics[width=0.95\textwidth]{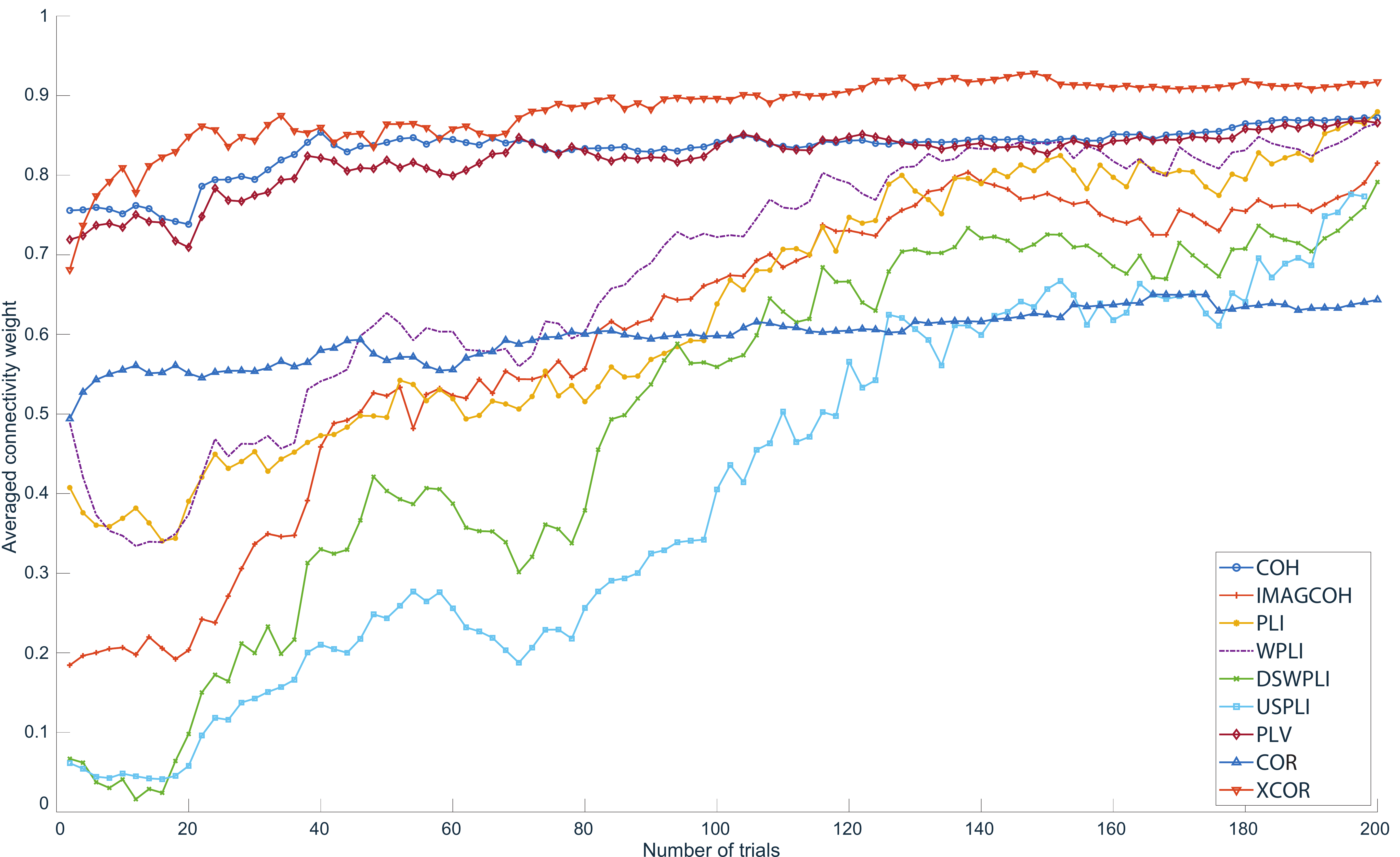}
\end{center}

\caption{The relationship between the final networks at 200 trials and networks calculated at different number of trials. The networks were estimated for all metrics and right-hand median nerve stimulation in an offline scenario.}
\label{fig:connectivity-experiment-eval}
\end{figure}

 Figure \ref{fig:connectivity-experiment-eval} presents the convergence of each metric with respect to the number of trials. As described before, for each metric the 20 strongest weighted edges at 200 trials were identified. The same edges and their averaged weights were computed for different numbers of trials subsequently. Metrics based on the imaginary part of the CSD stabilized slower than XCOR, COH, and PLV. The COR metric resulted in an almost stable average weight over all investigated numbers of trials. Again, please note that the convergence with larger number of trials does not reflect the correctness of the results; it merely reflects how quickly the detected stable network is established.
 
\subsubsection{Online functional connectivity evaluation} 

\begin{figure}

\begin{center}
a)\includegraphics[width=0.45\textwidth]{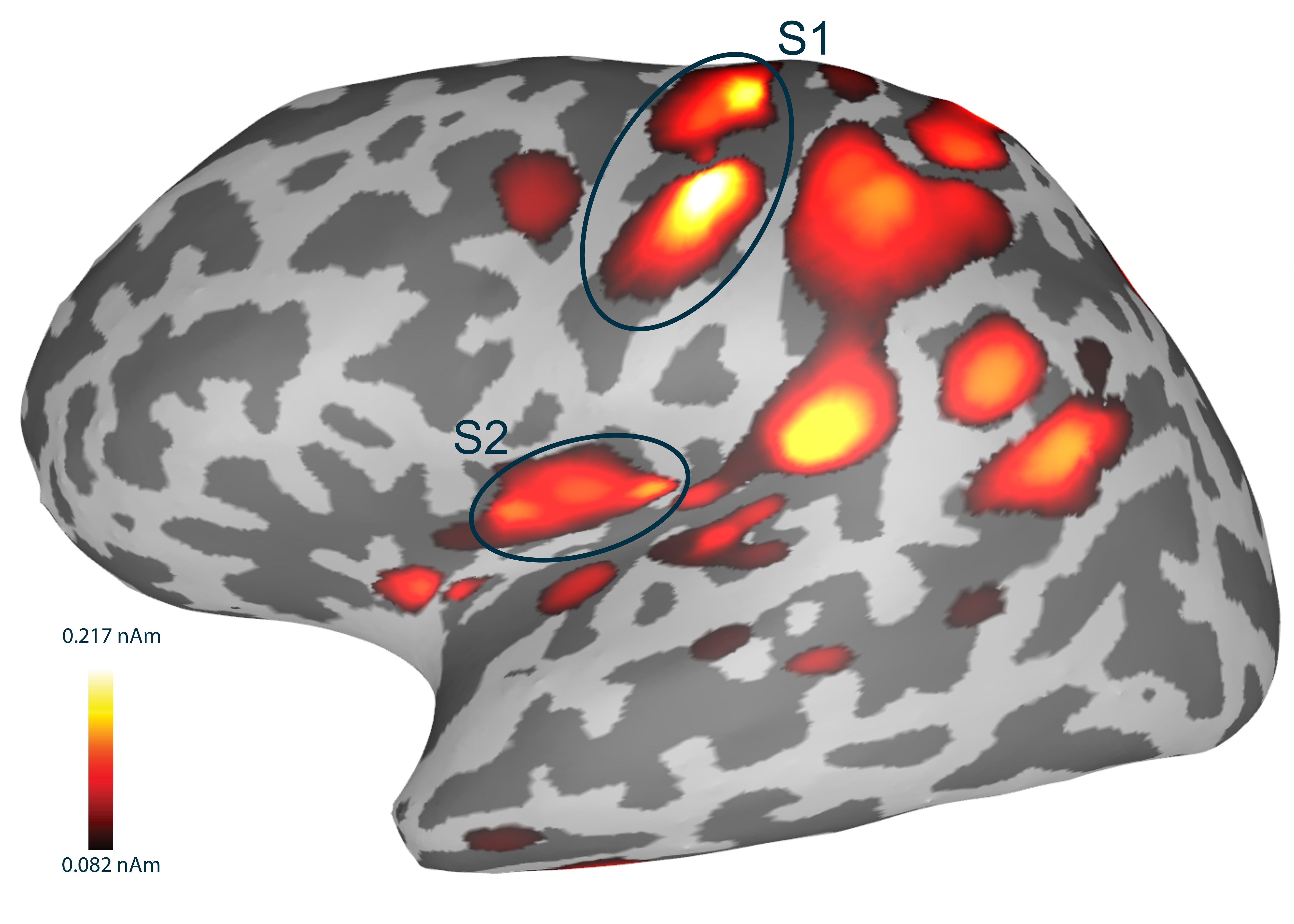} b)\includegraphics[width=0.45\textwidth]{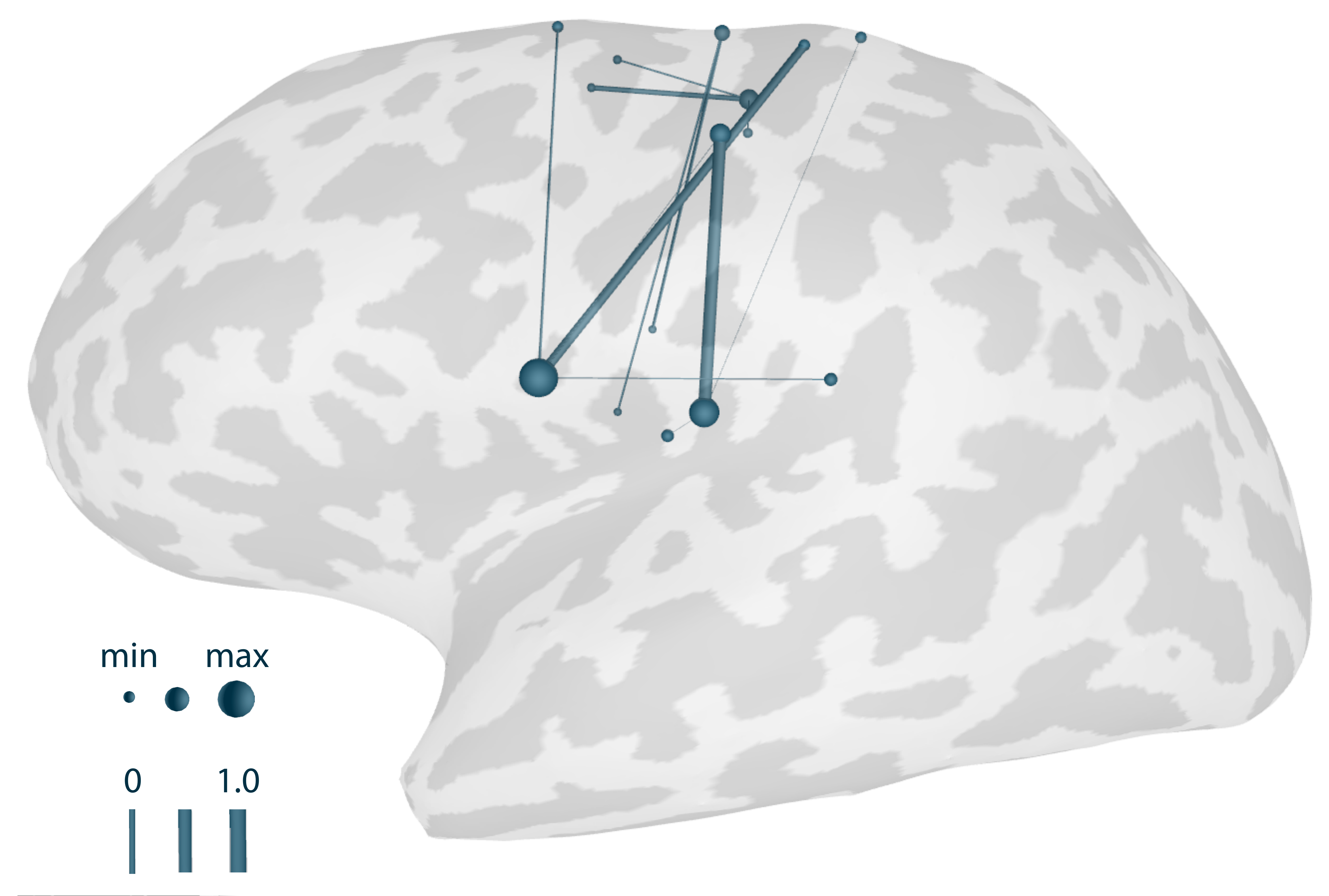}

\end{center}

\caption{a) One trial plotted as a 2D layout for median nerve stimulation. The result was produced in real-time, and the single trial RTC-MNE result for time point 35 ms is depicted. The cortical constrained interpolation method was used to plot the activity. The S1 and S2 areas are indicated by blue circles. The threshold was chosen to show only the strongest 75\% of activities. b) Visualization of an IMAGCOHY network estimated based on 200 good trials processed by the online connectivity pipeline. Only the edges representing the strongest 25\% of connections are plotted. Network nodes are plotted as spheres and edges are represented as tubes connecting the nodes. Edge strength and node degree are represented by their diameter. Please note that the nodes' sphere diameters are normalized by the node with the maximal value in the thresholded network.}
\label{fig:single-trial-eval}
\end{figure}

The source activity for the 265 sources was estimated for every incoming trial. Figure \ref{fig:single-trial-eval} shows the RTC-MNE result for a single trial. Although plotted for a single trial, an activation in both S1 and S2 was present. Please note that the absolute value of the source activation was plotted.

Based on the RTC-MNE result, the input data stream and its including data matrices had a dimension of 265x250. The functional connectivity for an all-to-all network consisting of 265 nodes was calculated. The intermediate data for each trial was stored, which meant the costly computations were done only once per trial. Frequency bins 0 to 50 were calculated and bins 18 to 30, corresponding to 18 to 30 Hz, were averaged to compute the edge weights. Each newly computed trial was added to the overall network. The visualization was updated every time a new network had been calculated. The different functional connectivity metrics were switched on the fly during the measurement session. The network was normalized based on the maximum edge weight. For visualization, the threshold was set to include the strongest 10\% of edges only (Figure \ref{fig:single-trial-eval}). A functional connection between the two regions of interest, S1 and S2, could be observed in real-time for the frequency band 18 to 30 Hz.
 
\begin{figure}

\begin{center}
\includegraphics[width=0.7\textwidth]{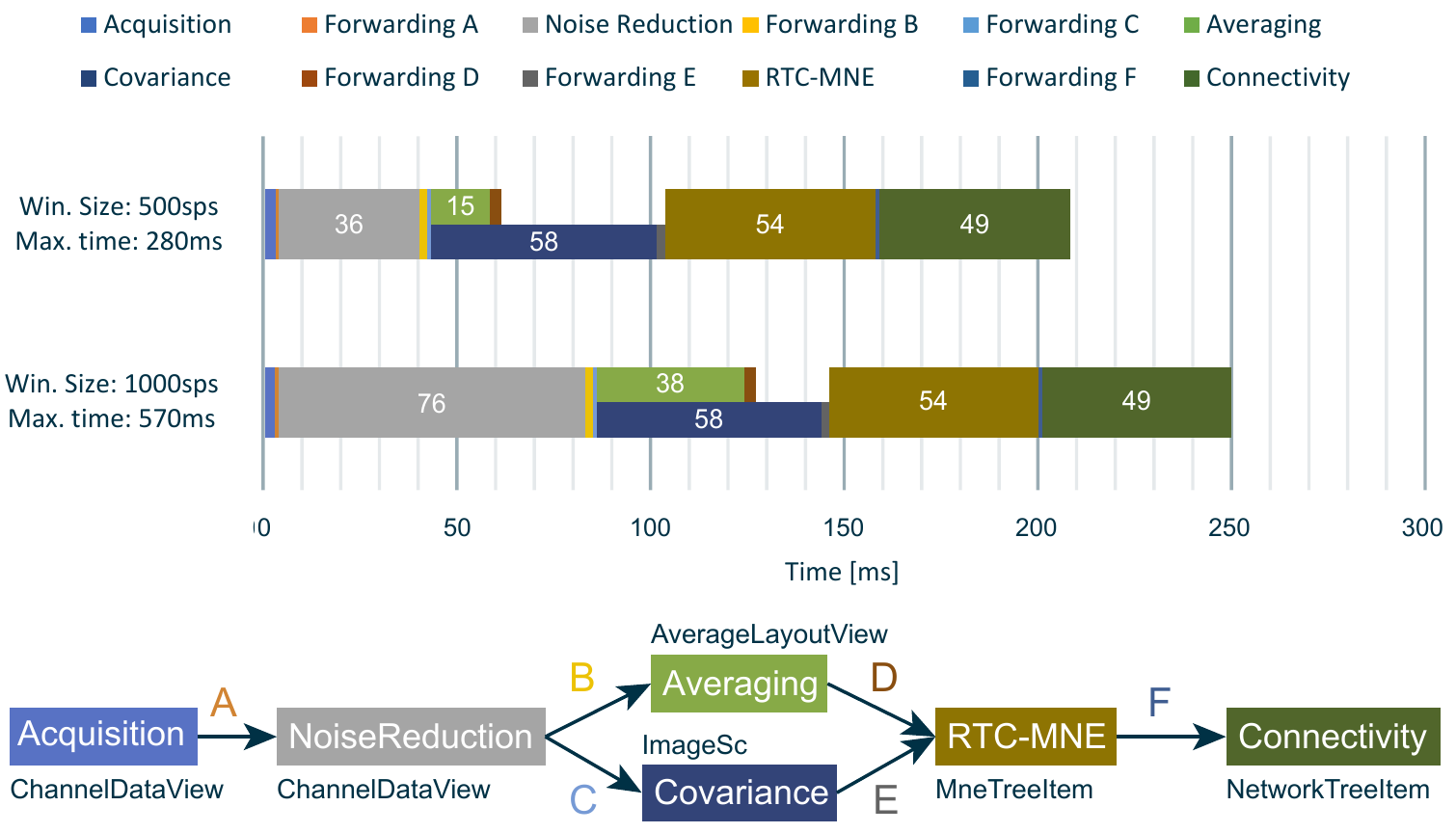}
\end{center}

\caption{The timing values in ms for each processing step corresponding to different block sizes and maximum allowed time limits. The lower plot indicates the different circular buffers between the plug-ins and the name of the corresponding visualization routine. }
\label{fig:single-trial-performance}
\end{figure} 

The performance of the presented online functional connectivity pipeline implemented in MNE Scan depends on several parameters including the sampling frequency, block size, number of measurement channels, and the processing as well as visualization complexity of the chosen pipeline.
To demonstrate these dependencies, the performance of the pipeline is presented in Figure \ref{fig:single-trial-performance}. The data were recorded at a sampling frequency of 1,792 Hz. Two scenarios with the same data set streamed with different block sizes (500 sp and 1,000 sp) were examined, resulting in different performance requirements for the overall pipeline timings. For example, choosing a block size of 500 sp at a 1,792 Hz sampling frequency results in a maximum allowed processing time of 280 ms. Not meeting this time requirement would introduce an overall delay and drag down the processing pipeline. The testing specifications were the same as in \citet{esch2018mne} except for the higher sampling frequency.

The timing values in Figure \ref{fig:single-trial-performance} show that for both window sizes and maximum allowed processing times, MNE Scan is indeed able to execute all processing steps online. This means the overall data pipeline is not slowed down by one of the individual processing steps. Please note that the averaging and covariance processing tasks are performed in parallel and not subsequent to each other. Their results are then forwarded independently to the RTC-MNE plug-in.
 
\section{Discussion}

The goal of this study was to present the efficient implementation of EEG/MEG functional connectivity metrics that support the computation of all-to-all connectivity networks in an easy-to-use toolbox for online EEG/MEG analysis, and demonstrate the applicability in realistic scenarios. Many previous studies restrict the number of nodes to a few ROIs in order to reduce the computational complexity or to suppress spurious connectivity. In the presented pipeline, the problem of computational burden is addressed by introducing the efficient calculation of functional connectivity metrics. Restricting the number of nodes requires prior knowledge about the process to be investigated and might result in missing out on important network nodes outside the selected ROIs. Therefore, it is desirable to include as many nodes as possible on the one hand, while keeping a meaningful and stable network on the other hand, which remains a challenge with current functional connectivity metrics. The median nerve stimulation inflicted in the proof-of-principle measurement produced a strong enough response and functional connectivity network to be estimated in the relatively large assumed all-to-all network.

In this work, all spectral metrics were implemented based on the CSD and PSD. PLV, PLI, USPLI, WPLI, and DSWPLI are pure phase measures, which can be calculated based only on the CSD of the analytic signal. IMAGCOHY and COH are most efficiently computed involving both CSD and PSD. From a performance standpoint and the fact that the metrics should be able to use intermediate data produced by other metrics, they should ideally share a large portion of their computational routines. Basing all spectral connectivity measures on CSD and PSD achieved that goal. This ensures an efficient and online capable implementation as well as convenient switching between metrics during an ongoing measurement session.
 
To date, only a few studies aimed to provide implementations of functional connectivity metrics that allow for an application in online scenarios. \citet{garcia2017efficient} describe tools to compute phase synchronization measures such as IMAGCOHY, PLV, PLI, and WPLI in real time. Compared to the present work, the authors chose a slightly different approach by first filtering the signal to the frequency band of choice and then generating the analytic signal via the Hilbert Transform. This enables the investigation of time-resolved functional connectivity by analyzing the instantaneous phase and amplitude. However, estimating these instantaneous components via the analytic signal only works reliably for very narrow bands or mono-component signals. Hence, a restriction of the signal to a narrow frequency band is required. In consequence, the subsequent functional connectivity estimation relies on the ability to design very narrow band temporal filters. Designing such filters is a non-trivial task and applying them involves another costly FFT convolution. Additionally, most of the analysis must be repeated if a different frequency band is selected. In an online processing pipeline doing trial resolved connectivity estimation, a change of the frequency band would mean that all trials would need to be processed again.
 \citet{garcia2017efficient} did not perform proof-of-principle measurements based on real human data. Instead, random data blocks were processed. Also, their tools were not integrated into a full online measurement setup including preprocessing and visualization.
 
\citet{hwang2007eeg,hwang2011eeg} implemented a functional connectivity EEG pipeline based on source-level data, and the source estimation was realized with MNE. The functional connectivity metric was chosen to be a simple COR and included 12 nodes. \citet{billinger2013single,billinger2015online} describe an Independent Component Analysis (ICA) based pipeline which is able to estimate effective connectivity on a small number of nodes. \citet{mullen2013real,mullen2015real} focus on a more advanced online pipeline including distributed source estimation, effective connectivity with a small number of nodes, and an online capable visualization including a GUI.

In comparison, the pipeline proposed in this work includes a device-independent acquisition with modular processing steps. These processing steps include temporal filtering, averaging, covariance estimation, source estimation, and connectivity analysis of large functional networks (\textgreater 200 nodes). Additionally, the result of each step can be visualized in 2D or 3D. To the best of our knowledge, the proposed pipeline is the first to combine device independence, online preprocessing, online distributed source-level reconstruction, online estimation and visualization of large functional connectivity networks, and GUI usage.

Except for \citet{garcia2017efficient}, no detailed performance values for online capable functional connectivity metrics were presented in prior publications. \citet{garcia2017efficient} evaluated the performance of the metrics based on a single trial with random data for different sample sizes and number of nodes. The hardware used to generate the performance values are different from the system in this work which provided 24 logical cores compared to 40 in \citet{garcia2017efficient}. The performance of the Connectivity library presented in this work outperforms the implementation presented in \citet{garcia2017efficient}, even though the computations were performed on a less powerful setup. However, this does not hold for window sizes smaller than 800 sp. This could be due to the extra code necessary to set up the data structures for multithreading in the implementation presented here. Please note that the multithreading setup is also executed when only one or a number of trials smaller than the number of CPU cores are processed.

The storage mode of the new Connectivity library enables the reuse of intermediate data generated in previous computations. This means that for every newly added trial only the evaluation of this single trial is necessary to update the connectivity network. However, it is important to mention that there are some code sections needed to be run in order to check if intermediate data from existing trials are available. This overhead needs to be considered when measuring the performance with the storage mode activated. With a large number of trials and analyzing multiple different metrics, the memory load corresponding to the stored intermediate data can be high. For this reason, the storage mode should only be used in demanding online setups together with hardware including a high memory capacity. In offline cases, the storage mode should be turned off in order to save memory. This will make the switching between metrics and recalculating metrics slower in favor of lower memory consumption.

The simulation-based source estimation results, both for the full and clustered source space, showed correct activity patterns in the selected ROIs. In the case of Label 2, the source activity spread to the temporal lobe, see Figure \ref{fig:reconstruction-simulation}, which can be explained by the curved cortical mantle. It is important to remember that the activity was visualized on an inflated surface. Whereas the visualization on the inflated surface gives the impression of a significant distance between the simulated source L2 and the found activation area on the temporal lobe, this distance is not present in the actual source space/cortex surface. The strongest active sources in both ROIs revealed the appropriate phase lagged time courses (Figure \ref{fig:reconstruction-simulation}).

The source reconstructed signals were used to validate the newly implemented metrics. All-to-all networks were computed between 243 nodes. The results show that the implemented metrics correctly identify functional connectivity between the two specified labels in the simulated data. Based on the fact that a 90$^{\circ}$ phase shift between the two labels' signal courses was introduced, it was to be expected that metrics based on the imaginary part of the CSD would be better suited to detect the signals' similarity. As shown by \citet{nolte2004identifying}, zero phase lag connectivity is discarded by IMAGCOHY and other metrics using the imaginary part of the CSD. This makes them more robust against the effects of volume conduction. This behavior is clearly reflected in the results presented in Figure \ref{fig:connectivity-simulation} for the IMAGCOHY, PLI, USPLI, and WPLI. Although XCOR is not a spectral metric it does account for temporal structure. That is why XCOR was able to detect the connectivity of the phase lagged signals as well. All other spectral metrics and COR were not able to identify a clear connection. These results confirm similar findings and previous discussions regarding the pitfalls of functional connectivity metrics \citep{bastos2016tutorial,palva2018ghost}.

Figure \ref{fig:connectivity-simulation-eval} shows that all spectral metrics and XCOR converge relatively quickly towards their final network structure. However, it can be seen that COH and PLV converge slightly faster than the other metrics. Please note that in this simulated case no additional sources were active except for the simulated ones. Consequently, this scenario is an idealized one with only the added noise interfering with the true connectivity.
 
The results of the proof-of-principle evaluation using realistic data show that meaningful functional connectivity networks can be reconstructed if a sufficient number of trials is considered. Figure \ref{fig:single-trial-eval} shows an IMAGCOHY-based network limited to the $\beta$-band (18 to 30 Hz), which is known to be linked to planning and processing of motor execution tasks. Both regions are assumed to be functionally connected during such processing. The actual network results and frequency bins per edge were averaged over this frequency band. The results align well with the ones produced in the simulated data, where only metrics based on the imaginary part of the CSD were able to estimate a meaningful connection by not being affected by the field spread.
 
The online results are comparable to the ones created offline prior to the online session (Figure \ref{fig:connectivity-experiment}). This is an expected result, since the offline and online results are computed using the exact same implementations with the only difference being that the latter was embedded in an online scenario with stringent time constraints. Figure \ref{fig:connectivity-experiment-eval} shows that metrics projected onto the imaginary plane take much longer to stabilize their final structure, but show meaningful connectivity between S1 and S2. Metrics based on both the real and imaginary part of the CSD need less trials to reach their final network structure, but are prone to volume conduction effects and do not show significant connectivity between S1 and S2. The difference in this final network generation speed can be explained by the fact that IMAGCOHY, PLI, USPLI, DSWPLI, and WPLI effectively discard information when being projected onto the imaginary plane. This is why they are more prone to noise and need more trials to produce clear results. On the contrary, metrics such as COH and PLV converge faster towards their final structure since they are not reduced in dimensionality. In contrast to the simulation study, COR and XCOR do not produce any plausible results when evaluating actual measurement data.

The overall timings of the final pipeline meet the online requirements as shown in Figure \ref{fig:single-trial-performance}. The presented timing values are based on operating systems not able to guarantee hard real-time requirements: Windows, macOS, and Linux hide specific thread prioritization and handling from the user. In order to meet stringent real-time requirements, meaning one can be certain that routines always take a predictable amount of time to run, Real-Time Operating Systems (RTOSs) can be employed. Examples of RTOSs are QNX (\url{http://blackberry.qnx.com/en/products/neutrino-rtos/index}) and RTX (\url{https://www.intervalzero.com/}). Since MNE-CPP and MNE Scan are solely based on Qt and Eigen, deploying them on a RTOS is feasible.

As reviewed and investigated in several prior publications, functional connectivity estimation is highly sensitive to a wide range of external factors and variations in pre-processing \citep{haufe2019simulation,palva2018ghost,sitaram2017closed,sakkalis2011review,barzegaran2017functional, schoffelen2009source,bastos2016tutorial, cho2015influence}. These include a low SNR, co-registration, artifacts, temporal filtering, the device to head transformation matrix, type of baseline correction, choice of source estimation method, and head model accuracy to name the most important ones. In the proposed pipeline, trial rather than time-resolved functional connectivity was computed in order to cope with the low SNR. Moreover, SSP and temporal filtering were employed to improve the SNR. A rather simple EOG-based artifact rejection was implemented, which excluded trials if a threshold was exceeded. It was already shown in a prior work that the RTC-MNE method is capable to cope with low SNR and eases computational complexity \citep{dinh2015real}. For this reason, the RTC-MNE method was chosen as the basis for the source-space functional connectivity evaluation.

The number of necessary trials to obtain stable functional connectivity networks observed in this study may seem very (or too) high for an online application. However, multiple studies proposed algorithms for brain-state classification based on functional connectivity estimates based on few or even single trials, e.g., for use in BCIs, though none of them actually applied their algorithm in an online scenario \citep{antonacci2019single,billinger2013single,rathee2017single,feng2020functional,shamsi2021early}. \citet{feng2020functional} demonstrated that sensor space functional connectivity estimates evaluated on different frequency bands with standard EEG preprocessing (re-referencing, temporal filtering) generates proper features for motor imagery (MI) classification, even outperforming a common spatial pattern (CSP) based approach. In a similar scenario and with standard EEG preprocessing, \citet{rathee2017single} additionally applied source reconstruction to avoid spurious connectivity due to volume conduction, and again showed that features derived from functional connectivity estimates can improve the classification accuracy in MI tasks in comparison to established methods such as CSP. Such approaches could be implemented in MNE Scan after the implementation of the respective classifiers, which is currently a work in progress.

\section{Conclusion}

The goal of this work was to integrate functional EEG/MEG connectivity analysis of large all-to-all networks into an online setting with a complete online pipeline including a GUI, which - to the best of our knowledge -  has not been accomplished so far. A proof-of-principle measurement showed that the implemented tools are indeed able to provide online functional connectivity analysis if necessary requirements are met.

The scope of this work did not include the development of new online capable metrics or necessary hardware. However, the results of this work confirm that new connectivity metrics and/or measurement hardware might be necessary in order to move towards true real-time connectivity estimation either on single-trial basis or to obtain an instantaneous result per sample. Nevertheless, the present work can be understood as a major step towards true real-time connectivity analysis as it demonstrated that efficient implementations of connectivity metrics enable the online processing of EEG/MEG data with existing and widely available computational resources.

The tools presented in this study allow to improve and ease access to online data monitoring for assessing data quality during rather than after the measurement enabling ruling out experimental problems early on. Moreover, they provide the foundation for the design of novel BCI applications and neurofeedback experiments based on functional connectivity networks. Integrating connectivity estimation into a neurofeedback setup is non-trivial and often requires single-trial or instantaneous rather than averaged results. Although this work primarily focused on the online estimation of evoked responses, spontaneous data can, from a technical point of view, be analyzed with the implemented tools, but estimating functional connectivity for spontaneous or resting-state data and related feature classification is the subject of ongoing research efforts \citep{billinger2013single,colclough2016reliable,garces2016quantifying,rathee2017single,feng2020functional,shamsi2021early}. For a reliable application of such pipelines based on MNE Scan in practice, further implementations and evaluations are necessary.

\section{Acknowledgements}
This work was supported by the Deutsche Forschungsgemeinschaft (DFG) under Grant
Ha2899/26-1 and by the Austrian Science Fund (FWF) under Grant I3790-B27 and
Grant P35949-B. This project received funding from the Free State of Thuringia (2018 IZN 004), co-financed by the European Union under the European Regional Development
Fund (ERDF).

\bibliography{connectivity}

\newpage
 
\appendix
\renewcommand{\thesection}{Supplementary Information \arabic{section}}    

\renewcommand{\thefigure}{SI~\arabic{figure}}
\setcounter{figure}{0}
\renewcommand{\thetable}{SI~\arabic{table}}
\setcounter{table}{0}

\section{Functional Connectivity Metrics}
\label{sec:metrics}

\subsection{Time Domain}
 For two signals in time domain, $x(t)$ and $y(t)$, we define 

 \subsubsection*{Correlation (COR)} 
 \begin{equation}
COR_{xy} = \frac{\sum_{t=1}^N (x(t)) - \bar{x}) (y(t) - \bar{y})}{\sqrt{\sum_{t=1}^N ( x(t) - \bar{x})^2 \sum_{t=1}^N(y(t) - \bar{y})^2}}
\end{equation}

\subsubsection*{Cross Correlation (XCOR)}

\begin{equation}
XCOR_{xy}(\tau) = \frac{\sum_{t=1}^{N-\tau} (x(t + \tau)) - \bar{x}) (y(t) - \bar{y})}{\sqrt{\sum_{t=1}^{N-\tau} ( x(t+\tau) - \bar{x})^2 \sum_{t=1}^N(y(t) - \bar{y})^2}}
\end{equation}
\begin{equation}
\{ XCOR_{xy} \in \mathbb{R} | -1 \leq COR_{xy} \leq 1 \}
\end{equation}

\subsection{Frequency Domain}
For two signals in the frequency domain, $x(\omega)$ and $y(\omega)$, with signal magnitudes $A_x(\omega)$ and $A_y(\omega)$ and phases $\Phi_x(\omega)$ and $\Phi_y(\omega)$, we define the power spectral density (PSD) and cross spectral density (CSD)

\begin{align}
PSD_x(\omega) &= A_x^2(\omega) \\
CSD_{xy}(\omega) &= x(\omega)y^*(\omega) \\
&= A_x(\omega) A_y(\omega) e^{i(\Phi_x(\omega) - \Phi_y(\omega))}
\end{align}

\subsubsection*{Coherency (COHY)}
\begin{align}
COHY_{xy}(\omega) &= \frac{\frac{1}{K} \sum_{k=1}^K A_x(\omega,k)A_y(\omega,k)e^{i(\Phi_x(\omega,k) - \Phi_y(\omega,k))}}{\sqrt{\left(\frac{1}{K} \sum_{k=1}^K A^2_x(\omega,k)\right)\left(\frac{1}{K} \sum_{k=1}^K A^2_y(\omega,k)\right)}} \\
&= \frac{\frac{1}{K} \sum_{k=1}^K CSD_{xy}(\omega,k)}{\sqrt{\left(\frac{1}{K} \sum_{k=1}^K PSD_y(\omega,k)\right)\left(\frac{1}{K} \sum_{k=1}^K PSD_y(\omega,k)\right)}}
\end{align}


\subsubsection*{Coherence (COH)}

\begin{equation}
COH_{xy}(\omega) = |COHY_{xy}(\omega)|
\end{equation}


\subsubsection*{Imaginary Part of Coherency (IMAGCOHY)}

\begin{equation}
IMAGCOHY_{xy}(\omega) = Im(COHY_{xy}(\omega))
\end{equation}


\subsubsection*{Phase Locking Value (PLV)}

\begin{align}
PLV_{xy} &= \left| \frac{1}{K} \sum_{k=1}^K e^{i(\Phi_x(\omega,k) - \Phi_y(\omega,k))} \right| \\
&=\left| \frac{1}{K} \sum_{k=1}^K\frac{CSD_{xy}(\omega)}{|CSD_{xy}(\omega)|} \right|
\end{align}


\subsubsection*{Phase Lag Index (PLI)}
\begin{equation}
PLI_{xy}(\omega) = \left| \frac{1}{K}\sum_{k=1}^K sign(Im(CSD_{xy}(\omega))) \right|
\end{equation}


\subsubsection*{Unbiase Squared PLI (USPLI)}
\begin{equation}
USPLI_{xy}(\omega) = \frac{K\,PLI_{xy}^2 - 1}{K - 1}
\end{equation}

\subsubsection*{Weighted PLI (WPLI)}
\begin{equation}
WPLI_{xy}(\omega) = \frac{|\frac{1}{K}\sum_{k=1}^K Im(CSD_{xy}(\omega))|}{\frac{1}{K}\sum_{k=1}^K |Im(CSD_{xy}(\omega)|}
\end{equation}

\subsubsection*{Debiased Squared WPLI (DSWPLI)}
\begin{equation}
DSWPLI_{xy}(\omega) = WPLI^2_{xy}(\omega)
\end{equation}

\newpage

\section{Figures}
\label{sec:supp-figures}
\begin{figure}

\begin{center}
\includegraphics[width=0.9\textwidth]{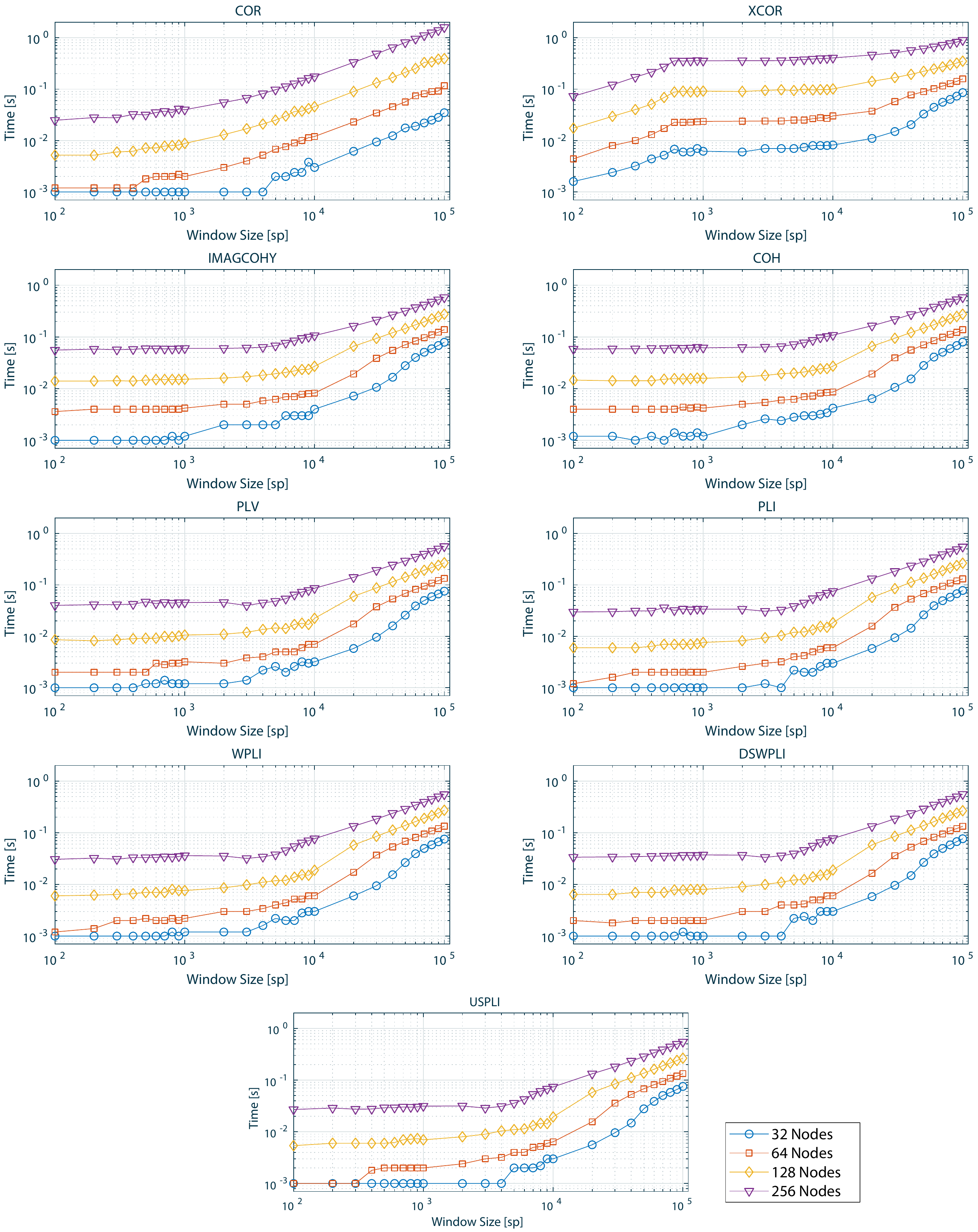}
\end{center}

\caption{Computational timing values in seconds for one trial and different window sizes in samples (sp) as well as number of computed nodes (y-axis in logarithmic scale). Computations were generated averaged over five repetitions.}
\label{fig:supp-timing1}
\end{figure}

\begin{figure}

\begin{center}
\includegraphics[width=0.9\textwidth]{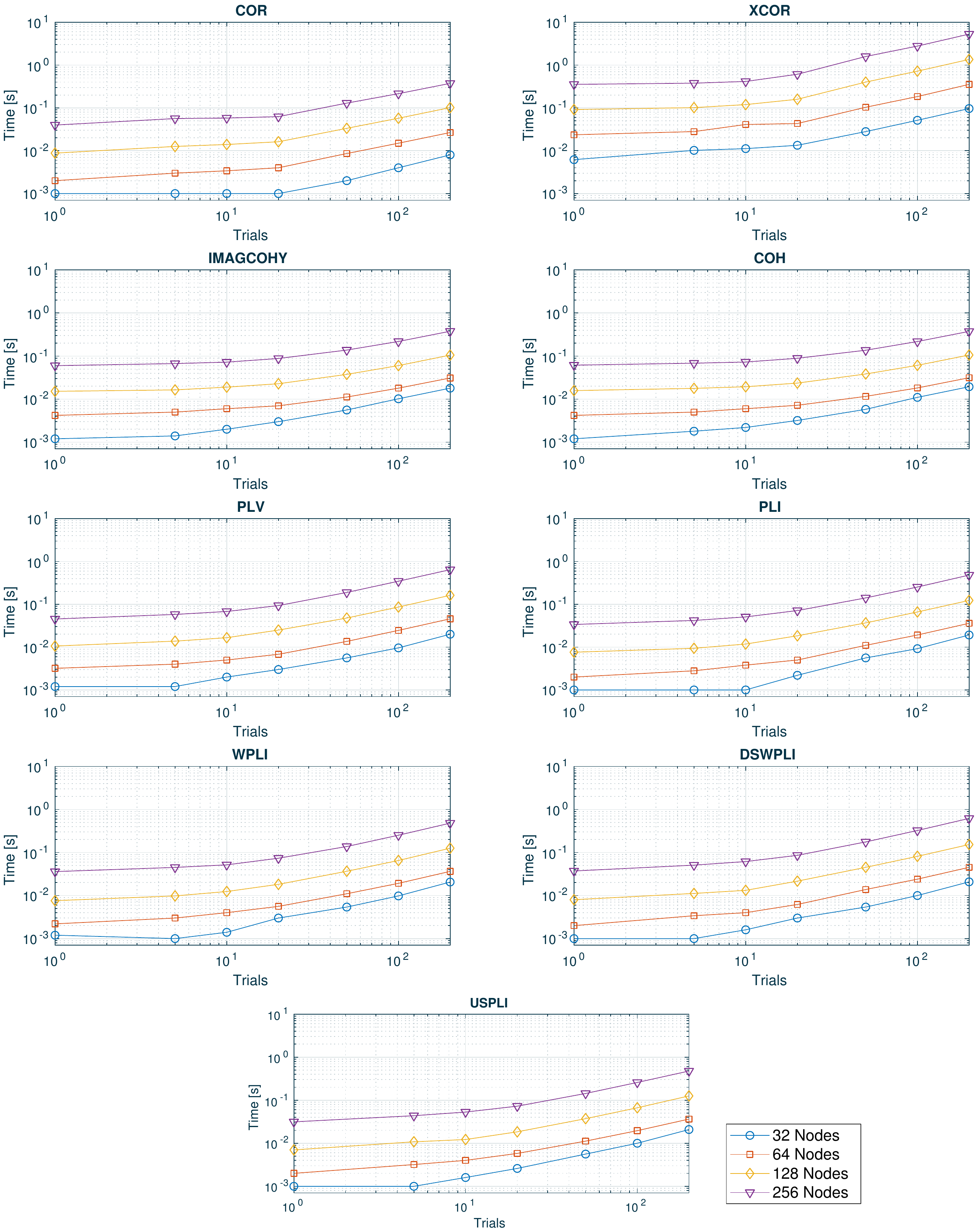}
\end{center}

\caption{Computational timing values in seconds for 1,000 sp and different number of trials as well as computed nodes (y-axis in logarithmic scale). Computations were generated averaged over five repetitions.}
\label{fig:supp-timing2}
\end{figure}

\begin{figure}

\begin{center}
\includegraphics[width=0.7\textwidth]{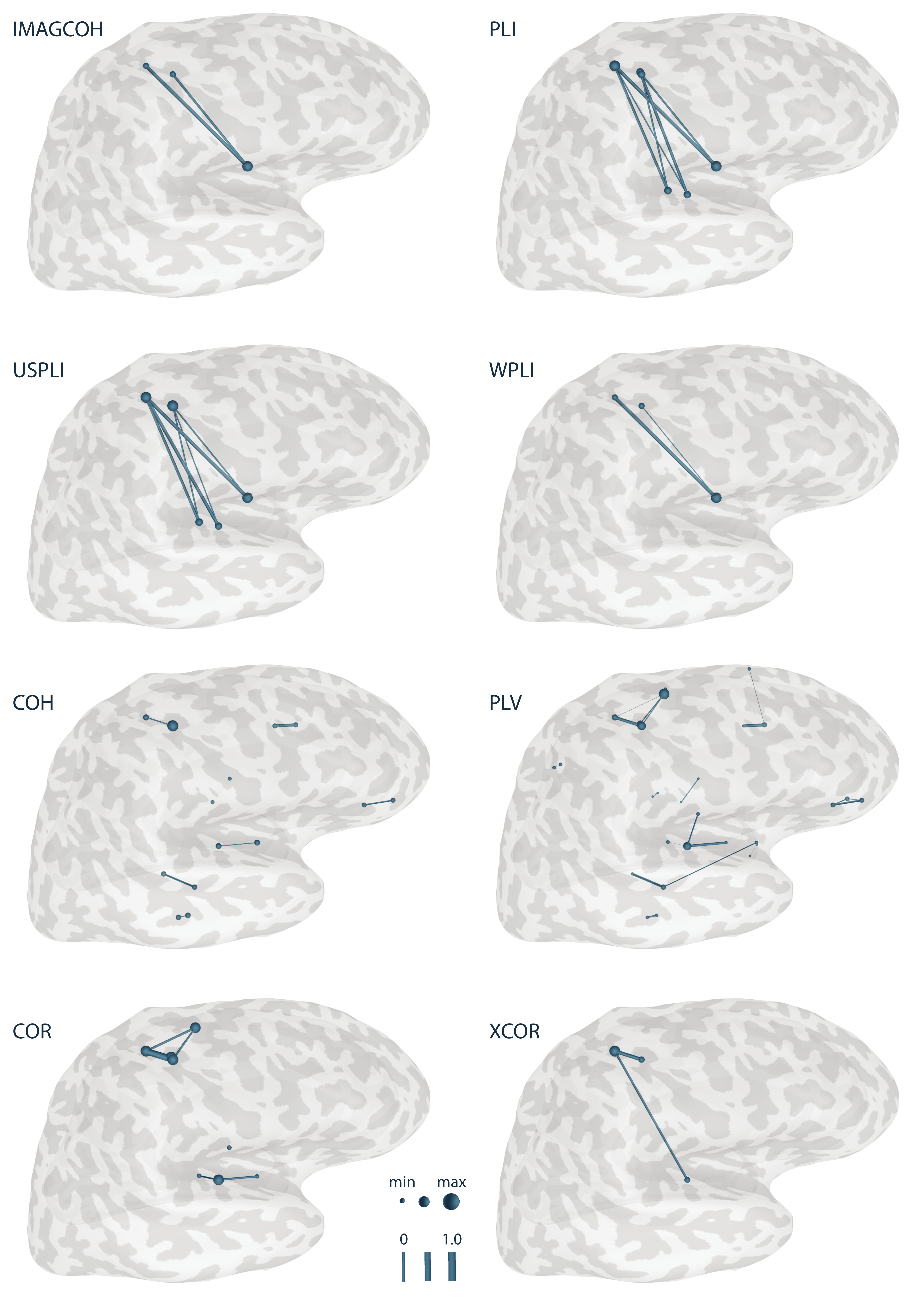}
\end{center}

\caption{Functional connectivity networks for different metrics based on simulated data. The RTC-MNE method was used to compute the source activity. The number of trials was 200. Network nodes are plotted as spheres and edges are represented as tubes connecting the nodes. Edge strength and node degree are represented by their diameter. Please note that the nodes' sphere diameters are normalized relative to the node with the maximal value in the thresholded network.}
\label{fig:supp-connectivity-simulation}
\end{figure}

\begin{figure}

\begin{center}
\includegraphics[width=0.45\textwidth]{fig_11_1} \includegraphics[width=0.45\textwidth]{fig_11_2}
\end{center}

\caption{Results for functional connectivity metrics implemented in the new Connectivity library based on right-hand median nerve stimulation. Results for 50 and 200 trials are presented. Only the edges representing the strongest 5\% of connections are plotted. Network nodes are plotted as spheres and edges are represented as tubes connecting the nodes. Edge strength and node degree are represented by their diameter. Please note that the nodes' sphere diameters are normalized relative to the node with the maximal value in the thresholded network.}
\label{fig:supp-connectivity-experiment}
\end{figure}
\end{document}